\documentclass[lettersize,journal]{IEEEtran}
\IEEEoverridecommandlockouts
\usepackage{cite}
\usepackage{amsmath,amssymb,amsfonts}
\usepackage{graphicx}
\usepackage{textcomp}
\usepackage{xcolor}
\usepackage{algorithm}
\usepackage{algpseudocode}

\usepackage{booktabs,array}
\usepackage{subfigure}
\usepackage{bm}
\usepackage{comment}
\usepackage{color}
\usepackage{stfloats}
\usepackage{makecell}
\usepackage{diagbox}

\begin{document}
	
	\title{CRLB Minimization for Multi-Target Localization in Segmented Waveguide-Enabled Pinching-Antenna ISAC Systems\\
	}
	
	\author{\IEEEauthorblockN{Yue Geng, Tee Hiang Cheng, Kah Chan Teh, \IEEEmembership{Senior Member, IEEE}, and Zhiguo Ding, \IEEEmembership{Fellow, IEEE}}
		\thanks{The work of the authors was supported by the School of Electrical and Electronic Engineering, Nanyang Technological University, Singapore. \emph{Corresponding author: Kah Chan Teh}.
			
			Yue Geng, Tee Hiang Cheng, Kah Chan Teh, and Zhiguo Ding are with the School of Electrical and Electronic Engineering, Nanyang Technological University, Singapore (Email: yue.geng@ntu.edu.sg; ethcheng@ntu.edu.sg; ekcteh@ntu.edu.sg; zhiguo.ding@ntu.edu.sg).
		}
	}
	
	\maketitle
	\begin{abstract}
		The segmented waveguide-enabled pinching-antenna system (SWAN) has been proposed to mitigate the inter-antenna radiation problem in uplink transmissions of conventional pinching-antenna (PA) systems. In this work, SWAN-assisted multi-user multi-target integrated sensing and communications (ISAC) is investigated, where a SWAN-enabled base station (BS) provides downlink communications for multiple communication users (CUs) and performs sensing for multiple targets. The dual-functional signals transmitted by the BS are radiated by the SWAN, and the echo signals reflected by the targets are captured by the SWAN and relayed to the BS for estimating the locations of the targets. We adopt the Cramér-Rao lower bound (CRLB) for target location estimation as the sensing metric, and formulate a CRLB minimization problem to evaluate the performance of the ISAC system. For joint PA position and beamforming optimization, we develop a Riemannian manifold optimization (RMO) method, in which a product variable representing both the beamforming and PA position variables is constrained on a constructed Riemannian product manifold (RPM). A penalty method with Riemannian Broyden–Fletcher–Goldfarb–Shanno (RBFGS) algorithm is then applied to update the product variable and obtain a feasible solution, enabling simultaneous and joint optimization of the variables. Simulation results demonstrate that the proposed RMO method achieves a lower CRLB than separate optimization approaches for SWAN. Compared with existing multi-waveguide PA-assisted ISAC systems, the SWAN optimized by the proposed RMO substantially reduces the required waveguide deployment length while incurring only a marginal CRLB performance loss under typical system settings.
	\end{abstract}
	
	\begin{IEEEkeywords}
		Segmented waveguide-enabled pinching-antenna system (SWAN), integrated sensing and communications (ISAC), Riemannian manifold optimization.
	\end{IEEEkeywords}
	
	\section{Introduction}
	Multiple-input multiple-output (MIMO) technology has played a central role in the evolution of wireless communications by providing substantial diversity, multiplexing, and array gains \cite{mimo}. To improve channel conditions in a flexible manner, reconfigurable antenna technologies have recently attracted increasing attention, enabling wireless systems to adapt their electromagnetic characteristics, such as radiation patterns and antenna locations \cite{ra}. Representative examples include reconfigurable intelligent surfaces (RISs) \cite{ris}, fluid antennas (FAs) \cite{fas}, and movable antennas (MAs) \cite{ma}, which exploit spatial reconfiguration to enhance wireless system performance and mitigate small-scale fading. However, the reconfigurability of these technologies is usually confined to limited apertures spanning only a few to several tens of wavelengths. As a result, although they are effective for small-scale channel adaptation, their capability to combat large-scale path loss remains limited. In addition, once deployed, modifying the number or arrangement of antennas in such systems often incurs considerable hardware cost and implementation complexity.
	
	To overcome these limitations, the pinching antenna system (PASS) has recently emerged as a promising large-scale reconfigurable antenna architecture \cite{pa1,pa2,pa3}. Originally proposed and prototyped by NTT DOCOMO, PASS employs low-attenuation dielectric waveguides as the primary transmission medium, whose propagation loss can be extremely small \cite{pa4}. In PASS, electromagnetic waves are guided through the dielectric waveguide and radiated into free space by attaching small separated dielectric particles, referred to as pinching antennas (PAs), at desired locations along the waveguide. By simply pinching or releasing these dielectric elements in a plug-and-play manner, PASS enables highly flexible and low-cost reconfiguration of both the number and positions of radiating points, thereby facilitating dynamic pinching beamforming \cite{pa5}. More importantly, since the waveguide can extend over meters or even larger-scale distances, PAs can be deployed close to users to establish strong line-of-sight (LoS) links, effectively reducing free-space path loss and mitigating blockage effects \cite{pa6}. In this sense, PASS provides a new way to transform path loss from an uncontrollable channel impairment into a partially programmable factor through flexible antenna placement. 
	
	Owing to these distinctive advantages, PASS has attracted growing interest as a scalable and practical alternative to conventional MIMO and other flexible-antenna technologies, particularly for future wireless communication and sensing systems. For example, in \cite{pa5}, the authors developed a physics-based hardware and signal model for PASS, and on that basis optimized transmit and pinching beamforming to demonstrate its substantial transmit-power reduction over conventional and massive MIMO systems. In \cite{pa6}, the energy efficiency maximization in a multi-user PASS was tackled by jointly optimizing the PA positions, transmit power, and time allocation under minimum rate constraints. The impact of in-waveguide path loss on robust beamforming and PA positioning was investigated by considering both lossy and lossless waveguides in \cite{pa7}. Besides, multi-waveguide-enabled PASS (MPASS) was studied in \cite{pa8} for multi-user communications, where an element-wise sequential optimization framework was proposed for obtaining favorable PA positions. These properties also make PASS a compelling candidate for sensing and integrated sensing and communications (ISAC) systems, where flexible antenna deployment and improved propagation conditions are highly desirable. For example, a PASS-based wireless sensing architecture was proposed in \cite{pa4sen1}, where the transmit waveform and PA positions were jointly optimized to improve multi-target sensing accuracy and robustness. Then, the application of PASS for ISAC was investigated in \cite{pa4isac1}, where a reinforcement learning-based optimization framework was proposed to enhance communication rate while satisfying sensing signal-to-noise ratio (SNR) requirements in PASS-assisted ISAC systems. An MPASS-assisted single-user single-target ISAC system was studied in \cite{pa4isac2}, focusing on the joint optimization of sensing SNR and communication rate. In \cite{pa4isac3}, this framework was extended to a multi-user single-target scenario, where beamforming and PA deployments were jointly optimized to minimize the Cramér-Rao lower bound (CRLB) of target parameter estimation under communication quality-of-service (QoS) constraints, demonstrating clear performance gains over benchmark schemes. In \cite{pa4isac4}, the CRLB for parameter estimation was also adopted as the sensing metric for PASS-enabled ISAC systems, where a PASS-equipped transmitter with a ULA-equipped receiver outperformed conventional ULA-based transceivers.
	
	Despite the flexibility, conventional PASS still has several limitations. In the uplink, when multiple PAs are deployed on the same waveguide, the received signal of one PA may propagate inside the waveguide and be re-radiated by other PAs, leading to the inter-antenna radiation (IAR) effect and making the uplink model difficult to characterize. In addition, for large-scale deployments, a long waveguide may introduce considerable in-waveguide propagation loss, which can offset the benefit of flexible PA placement. Long-waveguide architectures also suffer from limited maintainability, since fault localization and replacement are often costly. To address these issues, the segmented waveguide-enabled pinching antenna system (SWAN) was proposed, where a long waveguide is replaced by multiple short and isolated segments, each equipped with its own feed point. By confining propagation within each short segment, SWAN mitigates IAR, reduces in-waveguide loss, and improves maintainability. In \cite{swan1}, the SWAN architecture was developed to address the intractable uplink modeling and performance degradation issues of conventional long-waveguide PASS, where tractable uplink/downlink signal models, practical operating protocols, and PA placement methods were developed to maximize the received communication SNR. Multi-user uplink SWAN under different protocols was investigated in \cite{swan2}, where achievable sum-rate analyses and low-complexity PA placement methods were developed for both time-division multiple-access (TDMA) and non-orthogonal multiple-access (NOMA). The advantages of SWAN also make it a practical architecture for PASS-enabled sensing and ISAC systems. In \cite{swan3}, a downlink multi-user SWAN-assisted ISAC system was investigated, where a joint transmit beamforming and PA position optimization problem was formulated to maximize the sum rate under sensing constraints and solved via a reinforcement learning method. A SWAN-assisted ISAC system with separate transmit/receive SWANs was further studied in \cite{swan4}, where three segment-control protocols were introduced and the sensing limits as well as the sensing–communication trade-off were characterized and optimized.
	
	Although recent studies have demonstrated the potential of SWAN-assisted ISAC, their sensing designs have primarily focused on target illumination power or received sensing SNR, rather than target localization accuracy \cite{swan3,swan4}. However, maximizing the received sensing power does not necessarily guarantee accurate target localization, which also depends on the spatial sensitivity of target echo signals. This issue becomes more challenging when multiple targets are simultaneously sensed, as the PA positions must balance the localization accuracy of different targets while mitigating inter-user interference in communications, and is not well tackled in existing works. To address these challenges, we investigate a multi-user, multi-target SWAN-assisted ISAC system and formulate a joint PA positioning and beamforming optimization problem to minimize the target localization CRLB subject to communication requirements in this paper. To efficiently tackle this highly non-convex problem, we propose a Riemannian manifold optimization (RMO) method that exploits the geometric structure of the optimization variables for joint beamforming and PA position design. The main contributions are summarized as follows. 
	
	\begin{itemize}
		\item We investigate a SWAN-assisted ISAC system with multiple CUs and sensing targets. Unlike existing studies focusing on target illumination power or sensing SNR, we derive the CRLB for multi-target localization and formulate a CRLB minimization problem under communication constraints, thereby directly accounting for target localization accuracy in the joint ISAC design.
		
		\item To solve the resulting highly non-convex problem, we propose an RMO method that integrates the beamforming and PA position variables into a unified optimization variable on a Riemannian product manifold (RPM). By jointly and simultaneously optimizing these variables, the proposed method exploits their coupling and the geometric structure of the problem, avoiding the performance limitations of separate optimization.
		
		\item Simulation results demonstrate the effectiveness of the proposed SWAN architecture and RMO method. In particular, SWAN achieves lower CRLB than conventional MIMO and midpoint-based PA positioning schemes, while the proposed RMO method outperforms separate optimization approaches. Moreover, the RMO-optimized SWAN system substantially reduces the total deployment length of required waveguides compared with conventional multi-waveguide pinching-antenna systems, with only a marginal degradation in CRLB performance.
	\end{itemize}

	The rest of this paper is organized as follows. Section II presents the system model. Section III introduces the performance metrics and problem formulation. Section IV develops the proposed RMO method for joint beamforming and PA position optimization. Section V provides the simulation results, and Section VI concludes the paper.
	
	\textit{Notations:}  Scalars, vectors, and matrices are indicated as $a$, $\mathbf{a}$, and $\mathbf{A}$, respectively. $\mathbf{a}[i]$ and $\mathbf{A}[i,j]$ denote the $i$-th element of $\mathbf{a}$ and the element at the $i$-th row and $j$-th column of $\mathbf{A}$. $\mathbf{A}^\top$, $\mathbf{A}^{\mathrm{H}}$ and $\mathbf{A}^*$ indicate the transpose, conjugate transpose, and conjugate of $\mathbf{A}$, respectively. $\Re(\cdot)$ and $\Im(\cdot)$ denote taking the real part and imaginary part, respectively. $\mathbf{I}$ indicates identity matrix. $\odot$ denotes the element-wise multiplication. $\lvert \cdot \rvert$, $\lVert \cdot \rVert_2$, and $\lVert \cdot \rVert$ indicate modulus, 2-norm and Frobenius norm, respectively. $\operatorname{vec}(\cdot)$ and $\operatorname{Tr}(\cdot)$ are the vectorization and trace of a matrix, respectively. $\mathcal{U}(\cdot)$ denotes the uniform distribution.

	\section{System Model}
	
	In this section, we introduce the system model of the SWAN-assisted ISAC systems. As shown in Fig. 1, a BS provides downlink communications for $K_C$ CUs and performs sensing for $K_T$ targets. We denote the sets of the CUs and targets as $\mathcal{K}_C\triangleq\{1,\dots,K_C\}$ and $\mathcal{K}_T\triangleq\{1,\dots,K_T\}$, respectively. The locations of the $k$-th CU and target are $\mathbf{L}_{c,k}=(x_{c,k},y_{c,k},0),\forall k\in\mathcal{K}_C$ and $\mathbf{L}_{t,k}=(x_{t,k},y_{t,k},0),\forall k\in\mathcal{K}_T$, respectively. The CUs and targets are located in the rectangular service area $\mathcal{A}$ with side lengths $D_x$ and $D_y$. The BS transmits and receives signals via a SWAN system with height $d$. The considered system is modeled in a three-dimensional Cartesian coordinate system, where the spatial region is defined by $x\in [0,D_x]$, $y\in [-\frac{D_y}{2},\frac{D_y}{2}]$, and $z\in[0,d]$. The model details are introduced in the following subsections.
	
	\subsection{SWAN-Assisted Channel Models}
	As shown in Fig. 1, the SWAN system consists of $M$ transmit segmented waveguides (TSWs) and $M$ receive segmented waveguides (RSWs), which are arranged end-to-end without physical interconnection. The TSWs and RSWs are deployed at $y_{\rm tsw}=\frac{D_y}{2}$ and $y_{\rm rsw}=-\frac{D_y}{2}$, respectively, and are both aligned parallel to the $x$-axis. On each TSW, multiple transmit PAs (TPAs) could be activated to radiate signals to the CUs and targets, and we assume $N$ TPAs are activated on each TSW. On each RSW, one receive PA (RPA) is activated to receive the echo signals from the targets while avoiding the IAR effect. Each TSW or RSW has one feed point on its left endpoint, which is connected to the BS via a wired connection. Through the feed points, the signals transmitted by the BS are injected into the TSWs and radiated by the TPAs, or the echo signals reflected by the targets and received by the RPAs are extracted from the RSWs and relayed to the BS. The wired connections are assumed to be lossless since they incur negligible signal loss compared to the propagation losses within the segmented waveguides (SWs) and wireless channels. For simplicity, we assume that the lengths of the SWs are the same, and the length of each SW is $L$, which satisfies $D_x=LM$. For the $m$-th TSW, the location of the feed point is $\boldsymbol{\psi}_m^0\triangleq[\psi_m^0,y_{\rm tsw},d]^\top$, where $\psi_m^0=(m-1)L$. The location of the $n$-th PA deployed on it is denoted as 
	$\boldsymbol{\psi}^n_m\triangleq[\psi^n_m,y_{\rm tsw},d]^\top$. To ensure the TPAs are located on the dedicated TSW, the locations of the PAs should satisfy the constraints $\psi_m^0 \leq \psi^n_m \leq \psi_m^0+L$, and to avoid the coupling effect of the radiated signal, $\lvert \psi^n_m-\psi^{n'}_m\rvert \geq \frac{\lambda}{2}$ should be satisfied, where $\lambda$ is the signal wavelength. For the $m$-th RSW, the location of its feed point is $\boldsymbol{\phi}_m^0\triangleq[\phi_m^0,y_{\rm rsw},d]^\top$ with $\phi_m^0=(m-1)L$, and the location of the PA activated on it is $\boldsymbol{\phi}_m\triangleq[\phi_m,y_{\rm rsw},d]^\top$, which satisfies $\phi_m^0\leq \phi_m\leq \phi^0_m+L$. The vectors collecting the TPA and RPA position coordinates are denoted by $\boldsymbol{\psi}\triangleq\left[\psi_1^1,\dots,\psi_1^N,\psi_2^1,\dots,\psi_2^N,\dots,\psi_M^1,\dots,\psi_M^N\right]^\top\in\mathbb{R}^{NM}$ and $\boldsymbol{\phi}\triangleq\left[\phi_1,\dots,\phi_M\right]\in\mathbb{R}^M$, respectively.
	\begin{figure}[t]
		\centering{\includegraphics[width=1.0\columnwidth]{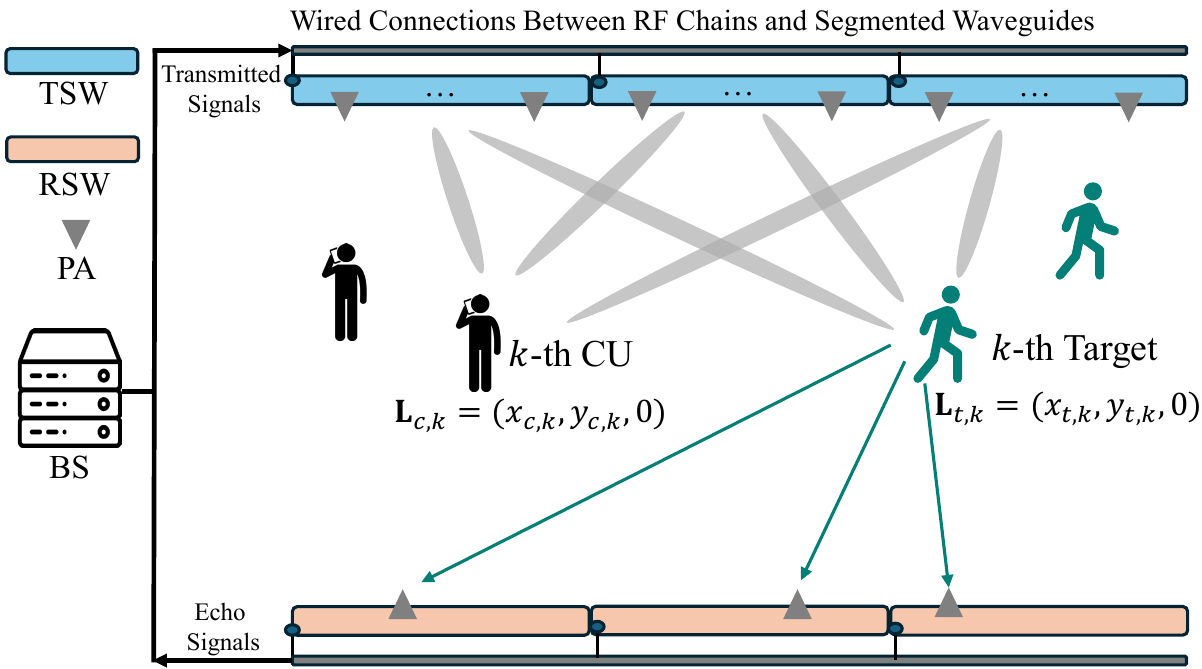}}
		\caption{Illustration of the proposed SWAN-assisted ISAC system.}
	\end{figure}
	\subsubsection{Communication Channel Model}
	For the downlink communication, the signals are transmitted via the in-waveguide channels, followed by the wireless PA-CU channels. Since the distances between the PAs deployed in the SWAN system are typically large compared with conventional antenna arrays, the CUs and targets are assumed to be located in the near-field region, and the spherical wave model is utilized to describe the propagation of the electromagnetic waves. Under the model, the channel coefficient between the $k$-th CU and the $n$-th TPA on the $m$-th TSW is given by
	\begin{align}
		g_{c,k}^{m,n}(\boldsymbol{\psi}_m^n,\mathbf{L}_{c,k})=\frac{\eta^{\frac{1}{2}}e^{{-\jmath k_c}D_{c,\psi}^{k,m,n}}}{D_{c,\psi}^{k,m,n}},
	\end{align}
	where $D_{c,\psi}^{k,m,n} = \lVert \mathbf{L}_{c,k}-\boldsymbol{\psi}_m^n\rVert=\sqrt{(x_{c,k}-\psi_m^n)^2+(y_{c,k}-y_{\rm tsw})^2+d^2}$, $\eta=\frac{\lambda^2}{16\pi^2}$, $\lambda$ is the free-space wavelength, $k_c=\frac{2\pi}{\lambda}$ is the wavenumber. Then, the overall channel between the PAs and the $k$-th CU is
	\begin{align}
		\mathbf{g}_{c,k}=[(\mathbf{g}^1_{c,k})^\top, \dots, (\mathbf{g}^M_{c,k})^\top]^\top\in\mathbb{C}^{NM}, \forall k\in\mathcal{K}_C,
	\end{align}
	where $\mathbf{g}^m_{c,k}=\left[g^{m,1}_{c,k},\dots,g^{m,N}_{c,k}\right]^\top\in\mathbb{C}^N$. We then adopt the equal power model, such that an equal proportion of the power allocated to a TSW is radiated by each TPA activated on it \cite{pa5}. For the in-waveguide channel, the propagation coefficient between the feed point and the $n$-th PA on the $m$-th TSW is 
	\begin{align}
		f_t^{m,n}(\boldsymbol{\psi}_m^n)=\frac{1}{\sqrt{N}}10^{-\frac{\kappa}{20}\lVert\boldsymbol{\psi}_m^n-\boldsymbol{\psi}_m^0\rVert}e^{-\jmath\frac{2\pi\lVert\boldsymbol{\psi}_m^n-\boldsymbol{\psi}_m^0\rVert}{\lambda_g}},
	\end{align}
	where $\kappa$ denotes the average attenuation factor along the waveguide, $\lambda_g=\frac{\lambda}{\eta_e}$ is the guided wavelength of the waveguide with effective refractive index $\eta_e$. In this work, we assume that each segmented waveguide operates in the fundamental single-guided-mode regime for tractable signal modeling, and the segment multiplexing operating protocol is implemented for the SWAN system, where each TSW is connected to its own dedicated RF chain, and a total of $M$ RF chains are connected to the BS. Thus, the overall in-waveguide channel between the BS and the TPAs is obtained as
	\begin{align}
		\mathbf{F}_t=\operatorname{Bdiag}(\mathbf{f}_t^1,\dots,\mathbf{f}_t^M)^\top\in\mathbb{C}^{M\times MN},
	\end{align}
	where $\mathbf{f}_t^m=[f_t^{m,1},\dots,f_t^{m,N}]^\top\in\mathbb{C}^{N}$. Then, the overall channel between the BS and the $k$-th CU is given as 
	\begin{align}
		\mathbf{h}_{c,k}=\mathbf{F}_t\mathbf{g}_{c,k}\in\mathbb{C}^M, \forall k\in\mathcal{K}_C,
	\end{align}
	where
	\begin{align}
		&\!\!\mathbf{h}_{c,k}[m](\mathbf{L}_{c,k},\boldsymbol{\psi}^n_m)\!=\!(\mathbf{f}_t^m)^\top\mathbf{g}_{c,k}\!\nonumber\\&=\!\!\sum_{n=1}^{N}\frac{\eta^{\frac{1}{2}}10^{-\frac{\kappa}{20}(\psi_m^n-\psi_m^0)}}{\sqrt{N}D_{c,\psi}^{k,m,n}}\cdot\! e^{-\jmath\left(k_cD_{c,\psi}^{k,m,n}+\frac{2\pi}{\lambda_g}(\psi_m^n\!-\psi_m^0)\right)}.
	\end{align}

	\subsubsection{Sensing Channel Model}
	Similar to the PA-CU channels, the wireless channel coefficient between the $k$-th target and the $n$-th PA on the $m$-th TSW is
	\begin{align}
		g_{t,k}^{m,n}(\boldsymbol{\psi}_m^n,\mathbf{L}_{t,k})=\frac{\eta^{\frac{1}{2}}e^{{-\jmath k_c}D_{t,\psi}^{k,m,n}}}{D_{t,\psi}^{k,m,n}},
	\end{align}
	where $D_{t,\psi}^{k,m,n} = \lVert \mathbf{L}_{t,k}-\boldsymbol{\psi}_m^n\rVert=\sqrt{(x_{t,k}-\psi_m^n)^2+(y_{t,k}-y_{\rm tsw})^2+d^2}$. Then the overall channel between the PAs and the $k$-th target is obtained as
	\begin{align}
		\mathbf{g}_{t,k}=[(\mathbf{g}^1_{t,k})^\top, \dots, (\mathbf{g}^M_{t,k})^\top]^\top\in\mathbb{C}^{NM}, \forall k\in\mathcal{K}_T,
	\end{align} 
	where $\mathbf{g}^m_{t,k}=\left[g^{m,1}_{t,k},\dots,g^{m,N}_{t,k}\right]^\top\in\mathbb{C}^N$. Since the communication and sensing channels share the same in-waveguide channel, the overall channel between the BS and the $k$-th target is
	\begin{align}
		\mathbf{h}_{t,k}=\mathbf{F}_t\mathbf{g}_{t,k}\in\mathbb{C}^M, \forall k\in\mathcal{K}_T,
	\end{align}
	where
	\begin{align}
		&\mathbf{h}_{t,k}[m](\mathbf{L}_{t,k},\boldsymbol{\psi}^n_m)=(\mathbf{f}_t^m)^\top\mathbf{g}_{t,k}\nonumber\\&=\!\!\sum_{n=1}^{N}\frac{\eta^{\frac{1}{2}}10^{-\frac{\kappa}{20}(\psi_m^n-\psi_m^0)}}{\sqrt{N}D_{t,\psi}^{k,m,n}}\cdot\!e^{-\jmath\left(k_cD_{t,\psi}^{k,m,n}+\frac{2\pi}{\lambda_g}(\psi_m^n\!-\psi_m^0)\right)}.
	\end{align}
	
	The signal reflected by the target can be captured by the RPA activated on the RSW and relayed to the BS via the waveguide and wired link. The wireless channel between the $k$-th target and the PA on the $m$-th RSW is given as
	\begin{align}
		g_{r,k}^{m}(\boldsymbol{\phi}_m,\mathbf{L}_{t,k})=\frac{\eta^{\frac{1}{2}}e^{{-\jmath k_c}D_{t,\phi}^{k,m}}}{D_{t,\phi}^{k,m}},
	\end{align}
	where $D_{t,\phi}^{k,m}=\lVert \mathbf{L}_{t,k}-\boldsymbol{\phi}_m\rVert=\sqrt{(x_{t,k}-\phi_m)^2+(y_{t,k}-y_{\rm rsw})^2+d^2}$. The in-waveguide channel between the $m$-th RPA and the corresponding feed point of the $m$-th RSW is
	\begin{align}
		f_r^{m}(\boldsymbol{\phi}_m)=10^{-\frac{\kappa}{20}\lVert\boldsymbol{\phi}_m-\boldsymbol{\phi}_m^0\rVert}e^{-\jmath\frac{2\pi\lVert\boldsymbol{\phi}_m-\boldsymbol{\phi}_m^0\rVert}{\lambda_g}}.
	\end{align}
	Then, the target-PA channel for the $k$-th target can be obtained as $\mathbf{h}_{r,k}=[h^1_{r,k}, \dots, h^M_{r,k}]^\top\in\mathbb{C}^{M}, \forall k\in\mathcal{K}_T$, where
	\begin{align}
		\mathbf{h}_{r,k}&[m](\mathbf{L}_{t,k},\boldsymbol{\phi}_m)= g_{r,k}^{m}(\mathbf{L}_{t,k},\boldsymbol{\phi}_m)\cdot f_r^{m}(\boldsymbol{\phi}_m)=\nonumber\\&\frac{\eta^{\frac{1}{2}}10^{-\frac{\kappa}{20}(\phi_m-\phi_m^0)}}{D_{t,\phi}^{k,m}}\cdot e^{-\jmath\left(k_cD_{t,\phi}^{k,m}+\!\frac{2\pi}{\lambda_g}(\phi_m-\phi_m^0)\right)}.
	\end{align}

	\subsection{Signal Model}
	In the considered ISAC system, the BS transmits $T$ signal symbol vectors during each ISAC period. We adopt a dual-functional transmission scheme in which communication signals are reused for sensing, while an additional dedicated sensing beam is introduced to enhance sensing performance \cite{signal}. The signal transmitted by the BS in the $t$-th time slot is designed as
	\begin{align}
		\mathbf{x}[t]=\mathbf{W}\mathbf{s}[t],
	\end{align}
	where $\mathbf{W}=[\mathbf{w}_{1},\dots,\mathbf{w}_{K_C},\mathbf{w}_S]\in\mathbb{C}^{M\times (K_C+1)}$ is the beamforming matrix, $\mathbf{w}_{k}$ is the beamforming vector for the $k$-th CU, and $\mathbf{w}_S$ is the dedicated sensing beamforming vector. $\mathbf{s}[t]=\left[s_1[t],\dots,s_{K_C}[t],s_S[t]\right]^{\top}\in\mathbb{C}^{K_C+1}$ is the signal symbol vector at the $t$-th time slot, where $s_k[t]$ contains the message for the $k$-th CU and $s_S[t]$ is the dedicated sensing symbol known to the BS. We assume that the signal symbols are zero mean, unit variance, and independent and identically distributed random variables, and are uncorrelated with each other, which leads to the condition $\mathbb{E}\{\mathbf{s}[t]\mathbf{s}^{\mathrm{H}}[t]\}=\mathbf{I}_{K_C+1}$. Besides, we assume that the sample number $T$ is sufficiently large in each ISAC period, such that the sample covariance matrix of $\mathbf{s}$ is approximately equal to its statistical covariance matrix, which is denoted as $\frac{1}{T}\sum_{t=1}^{T}\mathbf{s}[t]\mathbf{s}^{\mathrm{H}}[t]\approx \mathbf{I}_{K_C+1}$. 
	
	\subsubsection{Communication Signal Model} 
	For the downlink communication, the received signal of the $k$-th CU at the $t$-th time slot is given as
	\begin{align}
		\label{yc}
		y_k[t]&=\mathbf{h}^\top_{c,k}\mathbf{x}[t]+n_k[t]=\mathbf{h}^\top_{c,k}\mathbf{w}_ks_k[t]+\!\!\!\sum_{j\in\mathcal{K}_C\setminus\{k\}}\!\!\!\mathbf{h}^\top_{c,k}\mathbf{w}_js_j[t]\nonumber\\&\quad+\mathbf{h}^\top_{c,k}\mathbf{w}_S s_S[t]+n_k[t],
	\end{align}
	where $n_k[t]\sim\mathcal{CN}(0,\sigma_c^2)$ is the additive white Gaussian noise (AWGN) at the $k$-th CU receiver with noise variance $\sigma_c^2$. The dedicated sensing signal is treated as interference at the CUs.
	\subsubsection{Sensing Signal Model}
	For the radar sensing, the signal $\mathbf{x}[t]$ will be transmitted by the BS and radiated by the TPAs, propagate via the TPA-target channel, then be reflected back through the target-RPA channel. We assume that the direct transmitter-receiver leakage is sufficiently suppressed through self-interference cancellation, and therefore is omitted from the received signal. Thus, the echo signal received by the BS at the $t$-th time slot can be obtained as
	\begin{align}
		\mathbf{y}_r[t]=\sum_{k\in\mathcal{K}_T}\alpha_k \mathbf{h}_{r,k}\mathbf{h}^\top_{t,k}\mathbf{x}[t]+\mathbf{n}_s[t],
	\end{align}
	where $\alpha_k$ denotes the complex scattering coefficient of the $k$-th target, $\mathbf{n}_s[t]\sim\mathcal{CN}(0,\sigma_s^2\mathbf{I}_M)$ is the AWGN at the receiver of the BS, which contains the uncorrelated interference from the environment. Then, the target location estimation is performed by analyzing the received echo signals over the $T$ samples during one sensing period. Denote the echo channel of the $k$-th target as $\mathbf{H}_{t,k}=\alpha_k\mathbf{h}_{r,k}\mathbf{h}^\top_{t,k}$. By combining the $T$ samples, the received echo signal matrix can be denoted as
	\begin{align}
		\mathbf{Y}_r\!=\!\!\sum_{k\in\mathcal{K}_T}\mathbf{H}_{t,k}\mathbf{X}+\mathbf{N}_s\!\!=\!\!\sum_{k\in\mathcal{K}_T}\mathbf{H}_{t,k}\mathbf{W}\mathbf{S}+\mathbf{N}_s,
	\end{align}
	where $\mathbf{Y}_r=\left[\mathbf{y}_r[1],\dots,\mathbf{y}_r[T]\right]$, $\mathbf{X}=\left[\mathbf{x}[1],\dots,\mathbf{x}[T]\right]$, $\mathbf{S}=\left[\mathbf{s}[1],\dots,\mathbf{s}[T]\right]\in\mathbb{C}^{(K_C+1)\times T}$, and $\mathbf{N}_s=\left[\mathbf{n}_s[1],\dots,\mathbf{n}_s[T]\right]$. The parameter estimation is then implemented by analyzing the received echo signal $\mathbf{Y}_r$.

	\section{Performance Metrics and Problem Formulation}
	In this section, we introduce the performance metrics of the communication and sensing for the proposed SWAN-assisted ISAC systems and formulate the problem to test the performance of the systems. Specifically, we consider the location estimation of the targets as the sensing task, and we aim to minimize the CRLB of the location estimation while ensuring the constraints of the communication rate thresholds. 
	
	\subsection{Performance Metrics}
	\subsubsection{Communication Rate} 
	For the $k$-th CU, the achievable communication rate is $R_k=\log_2(1+\gamma_k)$, which is determined by the SINR obtained as
	\begin{align}
		\gamma_k=\frac{\lvert\mathbf{h}_{c,k}^\top\mathbf{w}_k\rvert^2}{\sum_{j\in\mathcal{K}_C\setminus\{k\}}\lvert\mathbf{h}_{c,k}^\top\mathbf{w}_j\rvert^2+\lvert\mathbf{h}_{c,k}^\top\mathbf{w}_S\rvert^2+\sigma_c^2}.
	\end{align}
	
	\subsubsection{CRLB of the Location Estimation}
	In this work, we consider the location estimation for the targets as the wireless sensing task. Since the targets are located in the $x-y$ plane, the locations are decided by the $x$-coordinates and $y$-coordinates, which are given as $\boldsymbol{x}=[x_{t,1},\dots,x_{t,K_T}]^\top$ and $\boldsymbol{y}=[y_{t,1},\dots,y_{t,K_T}]^\top$. Besides, the unknown target scattering coefficient parameters are denoted by $\boldsymbol{\alpha}_R=[\Re(\alpha_1),\dots,\Re(\alpha_{K_T})]^\top$ and $\boldsymbol{\alpha}_I=[\Im(\alpha_1),\dots,\Im(\alpha_{K_T})]^\top$.  Then, the total unknown parameter vector is 
	\begin{align}
		\boldsymbol{\theta} =\left[\boldsymbol{\xi}^\top,\boldsymbol{\alpha}^\top\right]^\top\in\mathbb{R}^{4K_T},
	\end{align}
	where $\boldsymbol{\xi}=[\boldsymbol{x}^\top,\boldsymbol{y}^\top]^\top\in\mathbb{R}^{2K_T}$ is the target parameter vector of interest to be estimated, and $\boldsymbol{\alpha} = [\boldsymbol{\alpha}_R^\top,\boldsymbol{\alpha}_I^\top]^\top\in\mathbb{R}^{2K_T}$ is the nuisance parameter. Considering the mean square error (MSE) between the estimated and true parameters as the sensing metric, the lower bound of the MSE obtained by any unbiased estimator is given as the CRLB. In this work, we utilize the CRLB as the sensing metric, and the development of specific estimators is considered for future work. 
	The received echo signal matrix in (17) is vectorized as
	\begin{align}
		\boldsymbol{y}_r
		&=\operatorname{vec}(\mathbf Y_r)
		=\sum_{k=1}^{K_T}
		\operatorname{vec}(\mathbf H_{t,k}\mathbf X)
		+\operatorname{vec}(\mathbf N_s).
	\end{align}
	Denoting
	$\boldsymbol{\Psi}=\sum_{k=1}^{K_T}
	\operatorname{vec}(\mathbf H_{t,k}\mathbf X)$,
	we have
	$\boldsymbol{y}_r\sim
	\mathcal{CN}(\boldsymbol{\Psi},
	\sigma_s^2\mathbf I_{MT})$. Note that the CRLB can be obtained as the diagonal elements of the CRLB matrix, which is the inverse of Fisher information matrix (FIM) \cite{fim}.
	The $(i,j)$-th element of the FIM associated
	with $\boldsymbol{\theta}$ is given by
	\begin{align}
		\mathbf F_{\theta}[i,j]
		=\frac{2}{\sigma_s^2}\Re\left\{
		\left(\frac{\partial\boldsymbol{\Psi}}
		{\partial\theta_i}\right)^H
		\frac{\partial\boldsymbol{\Psi}}
		{\partial\theta_j}
		\right\}.
	\end{align}
	By partitioning the FIM according to the
	target locations and the nuisance parameters,
	we obtain
	\begin{align}
		\mathbf F_{\theta}=
		\begin{bmatrix}
			\mathbf F_{\xi\xi} &
			\mathbf F_{\xi\alpha}\\
			\mathbf F_{\alpha\xi} &
			\mathbf F_{\alpha\alpha}
		\end{bmatrix} ,
	\end{align}
	where the expressions of the submatrices
	are provided in Appendix A. Then, by applying the Schur complement, the
	equivalent FIM for target localization
	is obtained as
	\begin{align}
		\mathbf F_{\rm eq}
		=\mathbf F_{\xi\xi}
		-\mathbf F_{\xi\alpha}
		\mathbf F_{\alpha\alpha}^{-1}
		\mathbf F_{\alpha\xi}.
	\end{align}
	Therefore, the CRLB for estimating the
	target locations is given by
	\begin{align}
		\mathrm{CRLB}_{\xi}
		=\operatorname{Tr}
		\left(\mathbf F_{\rm eq}^{-1}\right).
	\end{align}

	It should be noted that the calculation of $\operatorname{CRLB}_{\xi}$ requires the target locations $\boldsymbol{x}$ and $\boldsymbol{y}$ to be known. In this work, we consider the tracking stage of sensing, where the target locations and scattering coefficients estimated in previous sensing stages are utilized for CRLB minimization in the current stage. Accordingly, the proposed framework performs snapshot-based optimization at each sensing stage rather than jointly optimizing over dynamic target trajectories. For communication, the CSI can be acquired through advanced channel estimation techniques as investigated in recent studies on PASS \cite{CE}. We assume perfect CSI and target location information to focus on the proposed joint communication and sensing design in the following sections. Nevertheless, practical CSI and target location information may be imperfect due to estimation errors and user or target mobility. The impact of imperfect target location information on the proposed scheme will be evaluated through simulations in Section V.
	
	\subsection{Problem Formulation}
	To guarantee the QoS requirements of the CUs while enhancing the sensing performance of the ISAC, we aim to minimize the CRLB with respect to the target location estimation with the constraints of the communication rate thresholds by jointly optimizing the beamforming matrix and the positions of the TPAs and RPAs. The problem is formulated as
	\begin{subequations}
		\label{Q}
		\begin{IEEEeqnarray}{r,l}					
			$$\!\!\!\!\!\!\!\!\!\!\!\underset{\mathbf{W}, \boldsymbol{\psi}, \boldsymbol{\phi}}{\min}$$&{\ \operatorname{CRLB}_{\boldsymbol{\xi}}}\\
			$$\operatorname{s.t.}$$ &\ R_k\geq \Gamma_k,  \forall k\in\mathcal{K}_C\label{Cr}\\
			&\ \operatorname{Tr}(\mathbf{W}\mathbf{W}^{\mathrm{H}})= P_t,  \label{Cp}\\
			&\ \phi_m^0\!\leq\! \phi_m\!\leq\! \phi_m^0\!+\!L, \psi_m^0\!\leq\! \psi^n_m\!\leq\! \psi_m^0\!+\!L, \forall m,\!n, \label{Cregion} \\
			&\ \lvert \psi_m^n-\psi_m^{n'}\rvert\geq \frac{\lambda}{2}, \forall m, n\neq n', \label{Cd} 
		\end{IEEEeqnarray}
	\end{subequations}
	where \eqref{Cr} denotes the constraint of communication rate threshold, \eqref{Cp} indicates that the total transmit power budget of the system is $P_t$, \eqref{Cregion} restricts the TPAs and RPAs to be located on the dedicated SWs, \eqref{Cd} ensures the minimum distance between the TPAs to avoid the coupling effect. 
	
	We can observe that \eqref{Q} is hard to tackle due to the non-convex objective function and constraints. Existing works on SWAN-enabled wireless systems have typically optimized the PA positions and other variables separately. Particularly, the PA positions can be determined by projecting them onto the locations of the nearest user to minimize large-scale path loss \cite{swan1,swan4}. However, such schemes may no longer be effective in multi-user multi-target ISAC systems due to inter-user and inter-target interference. To obtain a more favorable solution, we develop the joint beamforming and PA position optimization for the SWAN-assisted multi-user multi-target ISAC in the following sections.
	
	\section{RMO for the Joint Beamforming and PA Position Optimization}
	
	In this section, we develop the RMO method to solve \eqref{Q}. Specifically, we first construct a Riemannian product manifold (RPM) as the solution domain. Then, the original problem is reformulated by integrating a penalty method with a smoothing technique. Finally, the variables are iteratively updated via the Riemannian Broyden–Fletcher–Goldfarb–Shanno (RBFGS) algorithm to obtain a feasible solution. 
	
	\subsection{Construction of the RPM}
	Riemannian manifold optimization generalizes classical unconstrained optimization from Euclidean spaces to smooth Riemannian manifolds (RMs) by directly exploiting the geometry of the search space. By equipping each tangent space of the manifold with a smoothly varying inner product, a Riemannian structure is established, which enables the definition of the Riemannian gradient and Hessian. This allows Euclidean optimization methods such as gradient descent and Newton’s method to be extended to manifold-constrained problems while naturally respecting the underlying geometric structure. Moreover, multiple RMs can be combined into an RPM to enable joint optimization of multiple variables \cite{rmo}. 
	
	The properties of the RPM motivate us to solve problem \eqref{Q} in this framework. First, \eqref{Cp} can be tackled by constraining the beamforming $\mathbf{W}$ on a Riemannian complex sphere manifold (CSM). Second, by constructing an RPM that incorporates the beamforming and PA position variables, a unified product variable can be constructed and optimized. The approach enables the PA positions and beamforming to be updated jointly and simultaneously, thereby better exploiting the coupling among the variables. To construct the RPM, by introducing an auxiliary variable $\tilde{\mathbf{W}}$ with projection $\mathbf{W} = \sqrt{P_t}\tilde{\mathbf{W}}$, the CSM is first given as
	\begin{flalign}	
		\label{MW}
		\mathcal{M}_{\tilde{\mathbf{W}}}=\{\tilde{\mathbf{W}}\in \mathbb{C}^{M\times (K_C+1)} \mid \operatorname{Tr}(\tilde{\mathbf{W}}\tilde{\mathbf{W}}^\mathrm{H}) = 1\}.
	\end{flalign}
	
	Observing \eqref{Cregion}, we note that the position variables $\boldsymbol{\psi}$ and $\boldsymbol{\phi}$ are not confined in RMs. To tackle \eqref{Cregion}, we introduce auxiliary variables $\tilde{\boldsymbol{\psi}}\triangleq\left[\tilde{\psi}_1^1,\dots,\tilde{\psi}_1^N,\tilde{\psi}_2^1,\dots,\tilde{\psi}_2^N,\dots,\tilde{\psi}_M^1,\dots,\tilde{\psi}_M^N\right]^\top\in\mathbb{R}^{NM}$ and $\tilde{\boldsymbol{\phi}}\triangleq\left[\tilde{\phi}_1,\dots,\tilde{\phi}_M\right]\in\mathbb{R}^M$. Then, we define projections
	\begin{align}
		\label{projpsi}
		\boldsymbol{\psi} &= \boldsymbol{p}(\tilde{\boldsymbol{\psi}})\nonumber\\&=\left[p_1\!\!\left(\tilde{\psi}_1^1\right),\dots,p_1\!\!\left(\tilde{\psi}_1^N\right),\dots,p_M\!\!\left(\tilde{\psi}_M^1\right),\dots,p_M\!\!\left(\tilde{\psi}_M^N\right)\right]
	\end{align}
	and
	\begin{align}
		\label{projphi}
		\boldsymbol{\phi}=\boldsymbol{q}(\tilde{\boldsymbol{\phi}})=\left[q_1\left(\tilde{\phi}_1\right),\dots,q_M\left(\tilde{\phi}_M\right)\right],
	\end{align}
	where $p_m(\tilde{\psi}_m^n)=(m-1)L+L\operatorname{sig}(\tilde{\psi}_m^n), \forall m,n$ and $q_m(\tilde{\phi}_m)=(m-1)L+L\operatorname{sig}(\tilde{\phi}_m), \forall m$. $\operatorname{sig}(x)=1/(1+e^{-\mu x})\in(0,1)$ denotes the sigmoid function, where $\mu$ is the scaling factor that controls the sigmoid slope and mitigates gradient saturation. The sigmoid mappings enforce the PA position constraints within the interior of each waveguide segment, while boundary positions can be approached arbitrarily closely. Note that $\tilde{\boldsymbol{\psi}}$ and $\tilde{\boldsymbol{\phi}}$ are confined in Euclidean real spaces, which are basic RMs \cite{rmo}. We then combine the beamforming matrix and the auxiliary PA position variables into a unified product variable, which is given as $\mathbf{X} = \left(\tilde{\mathbf{W}}, \tilde{\boldsymbol{\psi}}, \tilde{\boldsymbol{\phi}}\right)$. By integrating the RMs associated with each variable, an RPM can be constructed as
	\begin{align}	
		\label{M}
		\mathcal{M} \!=\!\{\mathbf{X}\!=\!(\tilde{\mathbf{W}}, \tilde{\boldsymbol{\psi}}, \tilde{\boldsymbol{\phi}})\! \mid\! \tilde{\mathbf{W}}\in\mathcal{M}_{\tilde{\mathbf{W}}}, \tilde{\boldsymbol{\psi}}\in\mathbb{R}^{NM},\tilde{\boldsymbol{\phi}}\in\mathbb{R}^M\}.
	\end{align}
	
	As introduced in \cite{rmo}, the RPM $\mathcal{M}$ can be locally linearized at each point via its corresponding tangent space. Specifically, for a point $\mathbf{X}$, the tangent space is given by
	\begin{align}	
		{\rm T}_{\mathbf{X}} \mathcal{M} =\bigl\{&\boldsymbol{\zeta}_{\mathbf{X}} = (\boldsymbol{\zeta}_{\tilde{\mathbf{W}}},\boldsymbol{\zeta}_{\tilde{\boldsymbol{\psi}}},\boldsymbol{\zeta}_{\tilde{\boldsymbol{\phi}}})\mid\nonumber\\& \boldsymbol{\zeta}_{\tilde{\mathbf{W}}}\in {\rm T}_{\tilde{\mathbf{W}}}\mathcal{M}_{\tilde{\mathbf{W}}},\boldsymbol{\zeta}_{\tilde{\boldsymbol{\psi}}}\in \mathbb{R}^{NM},\boldsymbol{\zeta}_{\tilde{\boldsymbol{\phi}}}\in \mathbb{R}^M\bigr\},
	\end{align}
	where the tangent space of the CSM is 
	\begin{flalign}	
		{\rm T}_{\tilde{\mathbf{W}}}\mathcal{M}_{\tilde{\mathbf{W}}} \!\!=\!\!\bigl\{\boldsymbol{\zeta}_{\tilde{\mathbf{W}}}\!\mid\! \boldsymbol{\zeta}_{\tilde{\mathbf{W}}}\!\in\!\mathbb{C}^{M\times(K_C+1)}, \Re\{\operatorname{Tr}({\tilde{\mathbf{W}}}^\mathrm{H}\boldsymbol{\zeta}_{\tilde{\mathbf{W}}})\}\!=\!0\bigr\}.
	\end{flalign}
	
	To define the inner product on the tangent space at $\mathbf{X}$, we adopt the Euclidean inner product as the Riemannian metric \cite{rmo}, which provides a simple yet effective way to measure lengths and angles in the local linear approximation and is expressed as
	\begin{align}	\langle\boldsymbol{\zeta}_{\mathbf{X}},\boldsymbol{\zeta}'_{\mathbf{X}}\rangle&=\langle(\boldsymbol{\zeta}_{\tilde{\mathbf{W}}},\boldsymbol{\zeta}_{\tilde{\boldsymbol{\psi}}},\boldsymbol{\zeta}_{\tilde{\boldsymbol{\phi}}}),(\boldsymbol{\zeta}'_{\tilde{\mathbf{W}}},\boldsymbol{\zeta}'_{\tilde{\boldsymbol{\psi}}},\boldsymbol{\zeta}'_{\tilde{\boldsymbol{\phi}}})\rangle\nonumber\\ &=  \Re\left(\operatorname{Tr}(\boldsymbol{\zeta}^\mathrm{H}_{\tilde{\mathbf{W}}}\boldsymbol{\zeta}'_{\tilde{\mathbf{W}}})\right)+  \boldsymbol{\zeta}^\top_{\tilde{\boldsymbol{\psi}}}\boldsymbol{\zeta}'_{\tilde{\boldsymbol{\psi}}} + \boldsymbol{\zeta}^\top_{\tilde{\boldsymbol{\phi}}}\boldsymbol{\zeta}'_{\tilde{\boldsymbol{\phi}}},
	\end{align}
	where $\boldsymbol{\zeta}_{\mathbf{X}}, \boldsymbol{\zeta}'_{\mathbf{X}} \in  {\rm T}_\mathbf{X}\mathcal{M}$, $\boldsymbol{\zeta}_{\tilde{\mathbf{W}}}, \boldsymbol{\zeta}'_{\tilde{\mathbf{W}}} \in  {\rm T}_{\tilde{\mathbf{W}}}\mathcal{M}_{\tilde{\mathbf{W}}}$, $\boldsymbol{\zeta}_{\tilde{\boldsymbol{\psi}}}, \boldsymbol{\zeta}'_{\tilde{\boldsymbol{\psi}}} \in  \mathbb{R}^{NM}$, and $\boldsymbol{\zeta}_{\tilde{\boldsymbol{\phi}}}, \boldsymbol{\zeta}'_{\tilde{\boldsymbol{\phi}}} \in  \mathbb{R}^M$. The norm of a tangent vector over the RPM is obtained as $\lVert \boldsymbol{\zeta}_{\mathbf{X}}\rVert = \sqrt{\langle\boldsymbol{\zeta}_{\mathbf{X}},\boldsymbol{\zeta}_{\mathbf{X}}\rangle}$.
	
	To enable consistent comparison of gradients and search directions, it is necessary to represent them within a common tangent space. To this end, a vector transport operation is employed, which maps tangent vectors from different tangent spaces to that at $\mathbf{X}$ while preserving their geometric properties \cite{rmo}. The operation over the RPM $\mathcal{M}$ is defined as
	\begin{align}
		\label{tra}
		\mathcal{T}_{\mathbf{X}}(\boldsymbol{\zeta}_{\mathbf{X}})= \bigl( \boldsymbol{\zeta}_{\tilde{\mathbf{W}}} - {\tilde{\mathbf{W}}}\Re\{\operatorname{Tr}({\tilde{\mathbf{W}}}^\mathrm{H}\boldsymbol{\zeta}_{\tilde{\mathbf{W}}})\},\boldsymbol{\zeta}_{\tilde{\boldsymbol{\psi}}},\boldsymbol{\zeta}_{\tilde{\boldsymbol{\phi}}}\bigr).
	\end{align}
	
	For a point $\mathbf{X} \in \mathcal{M}$, given a search direction $\mathbf{d}_{\mathbf{X}}$ and a step size $\alpha$, the retraction operation maps the updated point from the tangent space back onto the RPM $\mathcal{M}$, and is defined as
	\begin{align}
		\label{ret}
		\mathcal{R}_{\mathbf{X}}(\alpha \mathbf{d}_{\mathbf{X}}) =\bigl(\frac{{\tilde{\mathbf{W}}}+\alpha\mathbf{d}_{\tilde{\mathbf{W}}}}{\lVert {\tilde{\mathbf{W}}}+\alpha\mathbf{d}_{\tilde{\mathbf{W}}} \rVert}, \tilde{\boldsymbol{\psi}}+\alpha\mathbf{d}_{\tilde{\boldsymbol{\psi}}}, \tilde{\boldsymbol{\phi}}+\alpha\mathbf{d}_{\tilde{\boldsymbol{\phi}}} \bigr).
	\end{align}
	
	\subsection{Problem Reformulation}
	Exploiting the RPM $\mathcal{M}$, the inequality constraints \eqref{Cr} and \eqref{Cd} can be rewritten as
	\begin{align}
		h_k(\mathbf{X})=\Gamma_k-R_k \leq 0, \forall k\in\mathcal{K}_C
	\end{align} 
	and 
	\begin{align}
		h_{m,n,n'}(\mathbf{X})=\frac{\lambda^2}{4}-[ p_m(\tilde{\psi}_m^n)-p_m(\tilde{\psi}_m^{n'})]^2\leq0, \forall m, n\!\neq\! n',
	\end{align}
	respectively. By denoting the set of the minimum TPA distance constraints for all the TSWs as $\mathcal{L}= \{(m,n,n')\mid \forall m,n\neq n'\}$ and let $\mathcal{I}=\mathcal{L}\cup \mathcal{K}_C$, \eqref{Q} can be rewritten as
	\begin{subequations}
		\label{Q2}
		\begin{IEEEeqnarray}{r,l}			
			$$\underset{\mathbf{X}\in\mathcal{M}}{\min}$$&{\ f(\mathbf{X})=\operatorname{CRLB}_{\boldsymbol{\xi}}}\\
			$$\operatorname{s.t.}$$ &\ h_i(\mathbf{X})\leq 0,\forall i\in\mathcal{I}. \label{C_I}
		\end{IEEEeqnarray}
	\end{subequations}
	
	We then handle the inequality constraint \eqref{C_I} via a  penalty-based approach by augmenting the objective function with a weighted term, which penalizes constraint violations. We define the term as $\rho \sum_{i \in \mathcal{I}} \max\{0, h_i(\mathbf{X})\}$, where $\rho \geq 0$ is the penalty parameter. Under suitable regularity conditions, the nonsmooth exact penalty method can enforce constraint satisfaction with a finite penalty parameter in both Euclidean and Riemannian settings \cite{penalty,rep}. However, the resulting penalty term is nonsmooth and non-differentiable, making it difficult to optimize directly. To overcome this issue, we adopt a linear–quadratic approximation \cite{lq}, which provides a smooth approximation and is given by
	\begin{equation}
		\max\{0,x\}\approx\mathcal{P}(x,u)=
		\begin{cases}
			0& x\leq 0,\\
			\frac{x^2}{2u}& 0<x\leq u,\\
			x-\frac{u}{2}& x > u,
		\end{cases}
	\end{equation}
	where $u > 0$ is the smoothing parameter. In general, smaller values of $u$ lead to a closer approximation of the original nonsmooth penalty, while larger values improve smoothness and facilitate optimization. By incorporating this approximation, problem \eqref{Q2} can be reformulated as
	\begin{equation}
		\label{Qs}
		\underset{\mathbf{X}\in\mathcal{M}}{\min}\ g(\mathbf{X})=f(\mathbf{X})+\rho\sum_{i\in \mathcal{I}}\mathcal{P}(h_i(\mathbf{X}),u).
	\end{equation}
	
    Although the smoothed penalty function facilitates gradient-based optimization, it does not generally guarantee exact constraint satisfaction for finite penalty and smoothing parameters. Therefore, we aim to obtain a solution satisfying the inequality constraints within a prescribed tolerance by adjusting $\rho$ and $u$ and explicitly checking the constraint violations. However, although \eqref{Qs} is smooth over $\mathcal{M}$, it remains a non-convex problem and is therefore challenging to solve. To tackle these issues, we implement the RBFGS algorithm on $\mathcal{M}$ to solve \eqref{Qs} in the following subsections, which is further embedded within an outer iterative scheme to progressively adjust $\rho$ and $u$, thereby ensuring satisfaction of the inequality constraints.
	
	\subsection{RBFGS Algorithm Over the RPM}
	\subsubsection{Obtain the Riemannian Gradient}
	We first derive the Riemannian gradient of \eqref{Qs} with respect to $\mathbf{X}$ by projecting the corresponding Euclidean gradient onto the tangent space. The projection ensures that the resulting gradient lies on the manifold and respects its geometric structure. For any variable $\boldsymbol{\upsilon} \in \{\tilde{\mathbf{W}}, \tilde{\boldsymbol{\psi}}, \tilde{\boldsymbol{\phi}}\}$, the Euclidean gradient is computed as
	\begin{equation}
		\nabla_{\boldsymbol{\upsilon}} g(\mathbf{X})=\nabla_{\boldsymbol{\upsilon}}f(\mathbf{X})+\rho\sum_{i\in\mathcal{I}}\nabla_{\boldsymbol{\upsilon}} \mathcal{P}(h_i(\mathbf{X}),u),
	\end{equation}
	where
	\begin{equation}
		\nabla_{\boldsymbol{\upsilon}} \mathcal{P}(h_i(\mathbf{X}),u)\!\!=\!\!
		\begin{cases}
			\mathbf{0}& \!\!h_i(\mathbf{X})\leq 0,\\
			\frac{h_i(\mathbf{X})}{u} \nabla_{\boldsymbol{\upsilon}} h_i(\mathbf{X})&\!\! 0<h_i(\mathbf{X})\leq u,\\
			\nabla_{\boldsymbol{\upsilon}}h_i(\mathbf{X})& \!\!h_i(\mathbf{X}) > u.
		\end{cases}
	\end{equation}
	
	As the involved functions are smooth, the closed-form Euclidean gradients can be derived using standard matrix calculus \cite{CM}, which are provided in Appendix B. By aggregating these results, the overall Euclidean gradient of $g(\mathbf{X})$ with respect to $\mathbf{X}$ can be expressed as
	\begin{align}
		\nabla_{\mathbf{X}}g(\mathbf{X})=[\nabla_{{\tilde{\mathbf{W}}}^*}g(\mathbf{X}),\nabla_{\tilde{\boldsymbol{\psi}}}g(\mathbf{X}),\nabla_{\tilde{\boldsymbol{\phi}}}g(\mathbf{X})].
	\end{align}
	
	The Riemannian gradient can then be calculated based on the Euclidean gradient obtained. Specifically, the Riemannian gradient with respect to $\tilde{\mathbf{W}}$ is computed as the orthogonal projection of the Euclidean gradient onto the tangent space of the CSM, which is given as
	\begin{align}	
		\operatorname{grad}_{\tilde{\mathbf{W}}}g(\mathbf{X})\!=\!\nabla_{{\tilde{\mathbf{W}}}^*}g(\mathbf{X})\!-\! {\tilde{\mathbf{W}}}\Re\{\!\operatorname{Tr}\bigl(\!{\tilde{\mathbf{W}}}^\mathrm{H}\nabla_{{\tilde{\mathbf{W}}}^*}g(\mathbf{X})\!\bigr)\!\}.
	\end{align}
	
	Then for $\tilde{\boldsymbol{\psi}}$ and $\tilde{\boldsymbol{\phi}}$, we have $\operatorname{grad}_{\tilde{\boldsymbol{\psi}}}g(\mathbf{X})=\nabla_{\tilde{\boldsymbol{\psi}}}g(\mathbf{X})$ and $\operatorname{grad}_{\tilde{\boldsymbol{\phi}}}g(\mathbf{X})=\nabla_{\tilde{\boldsymbol{\phi}}}g(\mathbf{X})$. Following that, the Riemannian gradient with respect to $\mathbf{X}$ is obtained as
	\begin{align}
		\label{rgsr}
		\!\operatorname{grad}_{\mathbf{X}}g(\mathbf{X}) \!\!=\!\!\big[\!\operatorname{grad}_{{\tilde{\mathbf{W}}}}g(\mathbf{X}),\operatorname{grad}_{\tilde{\boldsymbol{\psi}}}g(\mathbf{X}),\operatorname{grad}_{\tilde{\boldsymbol{\phi}}}g(\mathbf{X})\!\big].
	\end{align}

	\subsubsection{Obtain Update Direction via RBFGS}
	
	Second-order methods can be more effective for non-convex optimization problems than first-order approaches, as they exploit curvature information to improve convergence. In principle, the Newton direction $\mathbf{d}_{\mathbf{X}}$ of $g(\mathbf{X})$ is obtained by solving the Newton equation $\operatorname{Hess}_{\mathbf{X}} g(\mathbf{X}) \mathbf{d}_{\mathbf{X}} = -\operatorname{grad}_{\mathbf{X}} g(\mathbf{X})$, where $\operatorname{Hess}(\cdot)$ denotes the Hessian operator \cite{rmo}. However, computing the Hessian is often computationally expensive, and it may be indefinite in non-convex settings, leading to instability. To address these issues, we employ the RBFGS algorithm to construct an efficient approximation of the inverse Hessian \cite{penalty}. By extending the BFGS algorithm from the Euclidean space to the RPM $\mathcal{M}$, the resulting search direction on $\mathcal{M}$ at the $l$-th iteration can be expressed as
	\begin{align}
		\label{d}
		\mathbf{d}_{\mathbf{X}}^l
		= (\mathbf{d}_{\tilde{\mathbf{W}}}^l,
		\mathbf{d}_{\tilde{\boldsymbol{\psi}}}^l,
		\mathbf{d}_{\tilde{\boldsymbol{\phi}}}^l)= -\mathbf{H}_{\mathbf{X}}^l
		\operatorname{grad}_{\mathbf{X}}g(\mathbf{X}^l),
	\end{align}
	where $\mathbf{H}_{\mathbf{X}}^l$ denotes the inverse Hessian approximation on the tangent space of the RPM at $\mathbf{X}^l$ \cite{rbfgs}. By operating on the entire product tangent space, the approximation can capture the coupling among the beamforming and PA position variables.
	
	To implement the RBFGS algorithm, we define intermediate variables
	$\mathbf{s}_{\mathbf{X}}^l =
	(\mathbf{s}_{\tilde{\mathbf{W}}}^l,
	\mathbf{s}_{\tilde{\boldsymbol{\psi}}}^l,
	\mathbf{s}_{\tilde{\boldsymbol{\phi}}}^l)$
	and
	$\mathbf{y}_{\mathbf{X}}^l =
	(\mathbf{y}_{\tilde{\mathbf{W}}}^l,
	\mathbf{y}_{\tilde{\boldsymbol{\psi}}}^l,
	\mathbf{y}_{\tilde{\boldsymbol{\phi}}}^l)$,
	which are given by
	\begin{align}
		\label{sy}
		\mathbf{s}_{\mathbf{X}}^l
		&= \mathcal{T}_{\mathbf{X}^{l+1}}
		(\alpha^l\mathbf{d}_{\mathbf{X}}^l),
		\nonumber\\
		\mathbf{y}_{\mathbf{X}}^l
		&= \operatorname{grad}_{\mathbf{X}}
		g(\mathbf{X}^{l+1})
		-\mathcal{T}_{\mathbf{X}^{l+1}}
		\bigl(\operatorname{grad}_{\mathbf{X}}
		g(\mathbf{X}^l)\bigr).
	\end{align}
	Both variables belong to the tangent space at $\mathbf{X}^{l+1}$. The inverse Hessian approximation is then updated as
	\begin{align}
		\label{bfgs}
		\mathbf{H}_{\mathbf{X}}^{l+1}
		=\mathbf{V}_{\mathbf{X}}^l
		\tilde{\mathbf{H}}_{\mathbf{X}}^l
		(\mathbf{V}_{\mathbf{X}}^l)^\mathrm{H}+\delta^l\mathbf{s}_{\mathbf{X}}^l
		(\mathbf{s}_{\mathbf{X}}^l)^\mathrm{H},
	\end{align}
	where $\tilde{\mathbf{H}}_{\mathbf{X}}^l$ denotes the inverse Hessian approximation transported to the tangent space at $\mathbf{X}^{l+1}$,
	$\mathbf{V}_{\mathbf{X}}^l
	=\mathbf{I}-\delta^l\mathbf{s}_{\mathbf{X}}^l
	(\mathbf{y}_{\mathbf{X}}^l)^\mathrm{H}$,
	and
	$\delta^l=
	1/\langle\mathbf{s}_{\mathbf{X}}^l,
	\mathbf{y}_{\mathbf{X}}^l\rangle$.
	Here, the rank-one operators and their adjoints are defined with respect to the Riemannian metric of the RPM.
	
	However, explicitly maintaining the inverse Hessian approximation in \eqref{bfgs} incurs considerable storage and computational overhead as the number of optimization variables increases. To alleviate this issue, we adopt a limited-memory strategy, which retains only a small set of information from recent iterations to construct the search direction \cite{lbfgs}. Specifically, we introduce a memory buffer $\mathbf{M}$ of size $S$, which stores up to $S$ sets of intermediate variables, denoted as
	$\mathbf{M}_i =
	(\mathbf{s}_{\mathbf{X}}^i,
	\mathbf{y}_{\mathbf{X}}^i,\delta^i)$,
	for $i=1,\ldots,m$, where $m\leq S$.
	The stored variables capture the curvature information required for approximating the inverse Hessian while keeping the storage and computational overhead manageable.
	
	Leveraging the recursive structure of \eqref{bfgs}, the inverse Hessian approximation can be expressed over $m$ updates as
	\begin{align}
		\label{lbfgs}
		&\mathbf{H}_{\mathbf{X}}^l
		=\bigl(\mathbf{V}_{\mathbf{X}}^{l-1}
		\cdots\mathbf{V}_{\mathbf{X}}^{l-m}\bigr)
		\tilde{\mathbf{H}}_{\mathbf{X}}^{l-m}
		\bigl(\mathbf{V}_{\mathbf{X}}^{l-1}
		\cdots\mathbf{V}_{\mathbf{X}}^{l-m}\bigr)^\mathrm{H}
		\nonumber\\
		&\ \ +\!\!\sum_{i=l-m}^{l-1}\!\!\delta^i
		\bigl(\mathbf{V}_{\mathbf{X}}^{l-1}
		\cdots\mathbf{V}_{\mathbf{X}}^{i+1}
		\mathbf{s}_{\mathbf{X}}^i\bigr)
		\!\!\times\!\!
		\bigl(\mathbf{V}_{\mathbf{X}}^{l-1}
		\cdots\mathbf{V}_{\mathbf{X}}^{i+1}
		\mathbf{s}_{\mathbf{X}}^i\bigr)^\mathrm{H},
	\end{align}
	where an empty product denotes the identity operator, and all tangent vectors and operators are understood to be appropriately transported to a common tangent space. The initial inverse Hessian approximation used in the limited-memory recursion is chosen as a scaled identity operator based on the most recent curvature pair \cite{lbfgs}. By combining \eqref{lbfgs} with \eqref{d}, the search direction $\mathbf{d}_{\mathbf{X}}^l$ can be efficiently computed via the standard two-loop recursion without explicitly constructing the inverse Hessian approximation \cite{lbfgs}. The overall procedure is summarized in \textbf{Algorithm 1}.
	
	\begin{algorithm}[t]
		\caption{RBFGS algorithm with limited memory}
		\begin{algorithmic}[1]
			\Require Initial direction
			$\mathbf{p}_{\mathbf{X}}^l
			=\operatorname{grad}_{\mathbf{X}}g(\mathbf{X}^l)$,
			$m$ ($m\leq S$) stored intermediate variable sets
			$\mathbf{M}_i=
			(\mathbf{s}_{\mathbf{X}}^i,
			\mathbf{y}_{\mathbf{X}}^i,\delta^i)$,
			$i=1,\ldots,m$, represented in the current tangent space.
			\For{$i=m:-1:1$}
			\State $\varrho^i
			=\delta^i
			\langle\mathbf{s}_{\mathbf{X}}^i,
			\mathbf{p}_{\mathbf{X}}^l\rangle$;
			\State $\mathbf{p}_{\mathbf{X}}^l
			=\mathbf{p}_{\mathbf{X}}^l
			-\varrho^i\mathbf{y}_{\mathbf{X}}^i$;
			\EndFor
			\If{$m>0$}
			\State $\mathbf{p}_{\mathbf{X}}^l
			=\frac{\langle\mathbf{s}_{\mathbf{X}}^m,
				\mathbf{y}_{\mathbf{X}}^m\rangle}
			{\langle\mathbf{y}_{\mathbf{X}}^m,
				\mathbf{y}_{\mathbf{X}}^m\rangle}
			\mathbf{p}_{\mathbf{X}}^l$;
			\EndIf
			\For{$i=1:m$}
			\State $\beta
			=\delta^i
			\langle\mathbf{y}_{\mathbf{X}}^i,
			\mathbf{p}_{\mathbf{X}}^l\rangle$;
			\State $\mathbf{p}_{\mathbf{X}}^l
			=\mathbf{p}_{\mathbf{X}}^l
			+(\varrho^i-\beta)\mathbf{s}_{\mathbf{X}}^i$;
			\EndFor
			\Ensure $\mathbf{d}_{\mathbf{X}}^l
			=-\mathbf{p}_{\mathbf{X}}^l$
		\end{algorithmic}
	\end{algorithm}
	
	After obtaining the update direction $\mathbf{d}_{\mathbf{X}}^l$ via \textbf{Algorithm 1}, an appropriate step size is determined using a line-search strategy to ensure sufficient descent of the penalized objective. The Armijo condition is given by
	\begin{equation}
		\label{Amijo}
		g(\mathbf{X}^{l+1})
		\leq g(\mathbf{X}^{l})
		+\sigma\gamma^n\tau_l
		\langle\operatorname{grad}_{\mathbf{X}}
		g(\mathbf{X}^{l}),
		\mathbf{d}_{\mathbf{X}}^{l}\rangle,
	\end{equation}
	where $\sigma,\gamma\in(0,1)$, and $\tau_l$ denotes the initial step size. By increasing $n$ until the Armijo condition \eqref{Amijo} is satisfied, the step size is determined as $\alpha^l=\gamma^n\tau_l$. For fixed penalty and smoothing parameters, this strategy ensures sufficient descent along a descent direction \cite{rmo}. The variable is then updated via the retraction operation as
	$\mathbf{X}^{l+1}
	=\mathcal{R}_{\mathbf{X}^l}
	(\alpha^l\mathbf{d}_{\mathbf{X}}^l)$.
	
	Subsequently, the intermediate variable set
	$(\mathbf{s}_{\mathbf{X}}^l,
	\mathbf{y}_{\mathbf{X}}^l,\delta^l)$
	is computed. To maintain a positive-definite inverse Hessian approximation \cite{rbfgs}, a cautious update strategy is employed to determine whether the new curvature pair should be stored for subsequent iterations. The cautious update condition is given by
	\begin{equation}
		\label{cautious}
		\langle\mathbf{s}_{\mathbf{X}}^l,
		\mathbf{y}_{\mathbf{X}}^l\rangle
		\geq 10^{-4}
		\langle\mathbf{s}_{\mathbf{X}}^l,
		\mathbf{s}_{\mathbf{X}}^l\rangle
		\lVert\operatorname{grad}_{\mathbf{X}}
		g(\mathbf{X}^l)\rVert.
	\end{equation}
	A curvature pair is stored only when \eqref{cautious} is satisfied and
	$\langle\mathbf{s}_{\mathbf{X}}^l,
	\mathbf{y}_{\mathbf{X}}^l\rangle>0$,
	so that $\delta^l$ is well defined.
	
	It should be noted that, when implementing \textbf{Algorithm 1} in subsequent iterations, the stored intermediate variables
	$(\mathbf{s}_{\mathbf{X}}^i,
	\mathbf{y}_{\mathbf{X}}^i,\delta^i)$
	must be represented in the tangent space at the current point through vector transport, with the curvature coefficients updated accordingly, so that all quantities remain geometrically consistent on the RPM. In addition, due to the limited memory size, once the memory buffer is full, the oldest intermediate variable set is removed to store the newly generated one.

	\subsection{Summary of the RMO Method}
	The effectiveness of solving \eqref{Qs} to obtain a feasible solution to \eqref{Q2} depends critically on the choice of the penalty parameter $\rho$ and the smoothing parameter $u$. In general, the inequality constraint \eqref{C_I} is more likely to be satisfied when $\rho$ is sufficiently large and $u$ is sufficiently small \cite{penalty}. Nevertheless, choosing an excessively large $\rho$ may lead to slow convergence and numerical instability. To balance constraint enforcement and optimization efficiency, we adopt an exact penalty strategy that iteratively updates $\rho$ and $u$ while solving \eqref{Qs} \cite{rep}. Specifically, whenever the solution to \eqref{Qs} does not satisfy \eqref{C_I}, the penalty parameter is increased according to $\rho = \theta_\rho \rho$, where $\theta_\rho > 1$. Meanwhile, the smoothing parameter is reduced as $u = \max\{u_{\min}, \theta_u u\}$, where $\theta_u \in (0,1)$ and $u_{\min}$ denotes a predefined lower bound. In this way, the approximation becomes progressively closer to the original nonsmooth penalty while maintaining numerical tractability. Accordingly, a feasible solution to \eqref{Q2} can be obtained by repeatedly solving \eqref{Qs} and updating $\rho$ and $u$ until the constraint \eqref{C_I} is met.

	\begin{algorithm}[t]
		\caption{The proposed RMO method for solving \eqref{Q2}}
		\begin{algorithmic}[1]
			\Require $l=0$, $\rho^0>0$, $u^0>u_{\min}>0$,
			$\theta_\rho>1$, $\theta_u\in(0,1)$,
			convergence threshold $\tau$, $S$, $I_{\max}$, $L_{\max}$.
			\State Initialize $\mathbf{X}_{\mathrm{out}}^0$
			via the ZF-based EWSO method.		
			\Repeat
			\State $i=0$,
			$\mathbf{X}^0=\mathbf{X}_{\mathrm{out}}^l$,
			$\mathbf{M}=\emptyset$, $m=0$;	
			\State Set $g(\mathbf{X})$ using $\rho^l$ and $u^l$;
			\Repeat
			\State Obtain
			$\operatorname{grad}_{\mathbf{X}}
			g(\mathbf{X}^i)$ by \eqref{rgsr};	
			\State Obtain $\mathbf{d}_{\mathbf{X}}^i$
			by \textbf{Algorithm 1};	
			\State Obtain $\alpha^i$
			via line-search strategy \eqref{Amijo};	
			\State Update
			$\mathbf{X}^{i+1}=
			\mathcal{R}_{\mathbf{X}^i}
			(\alpha^i\mathbf{d}_{\mathbf{X}}^i)$;	
			\State \parbox[t]{0.85\linewidth}{
				Compute $\mathbf{s}_{\mathbf{X}}^i$
				and $\mathbf{y}_{\mathbf{X}}^i$ via \eqref{sy};
			}			
			\State \parbox[t]{0.85\linewidth}{
				Transport the stored curvature pairs in
				$\mathbf{M}$ to
				$\mathrm{T}_{\mathbf{X}^{i+1}}\mathcal{M}$
				and update their curvature coefficients;
			}			
			\If{\eqref{cautious} is satisfied}	
			\State Compute
			$\delta^i=
			1/\langle\mathbf{s}_{\mathbf{X}}^i,
			\mathbf{y}_{\mathbf{X}}^i\rangle$;
			
			\If{$m<S$}
			\State $m=m+1$;
			\Else
			\State \parbox[t]{0.75\linewidth}{
				Remove $\mathbf{M}_1$ and set
				$\mathbf{M}_j=\mathbf{M}_{j+1}$,
				$j=1,\ldots,S-1$;
			}
			\EndIf
			
			\State Store
			$\mathbf{M}_m=
			(\mathbf{s}_{\mathbf{X}}^i,
			\mathbf{y}_{\mathbf{X}}^i,\delta^i)$;
			
			\EndIf
			
			\State $i=i+1$;
			
			\Until{$
				\lVert\mathbf{X}^i-\mathbf{X}^{i-1}\rVert
				<\tau
				$ or $i\geq I_{\max}$}
			
			\State
			$\mathbf{X}_{\mathrm{out}}^{l+1}
			=\mathbf{X}^i$;
			
			\If{$h_j(\mathbf{X}_{\mathrm{out}}^{l+1})
				>\tau$, $\exists j\in\mathcal{I}$}
			\State
			$\rho^{l+1}=\theta_\rho\rho^l$;
			\Else
			\State
			$\rho^{l+1}=\rho^l$;
			\EndIf
			
			\State
			$u^{l+1}
			=\max\{u_{\min},\theta_u u^l\}$;
			
			\State $l=l+1$;
			
			\Until{$
				\bigl(
				\lVert\mathbf{X}_{\mathrm{out}}^l
				-\mathbf{X}_{\mathrm{out}}^{l-1}\rVert
				<\tau
				\text{ and }
				h_j(\mathbf{X}_{\mathrm{out}}^l)\leq\tau,
				\forall j\in\mathcal{I}
				\text{ and }
				u^l\leq u_{\min}
				\bigr)
				$ or $l\geq L_{\max}$}
			
			\Ensure
			$\mathbf{X}_{\mathrm{out}}^l=
			(\tilde{\mathbf{W}}^l,
			\tilde{\boldsymbol{\psi}}^l,
			\tilde{\boldsymbol{\phi}}^l)$,
			$\mathbf{W}=\sqrt{P_t}\tilde{\mathbf{W}}^l$,
			$\boldsymbol{\psi}
			=\boldsymbol{p}(\tilde{\boldsymbol{\psi}}^l)$,
			$\boldsymbol{\phi}
			=\boldsymbol{q}(\tilde{\boldsymbol{\phi}}^l)$.
		\end{algorithmic}
	\end{algorithm}

	Since the proposed RMO method may converge to a suboptimal local minimum, a zero-forcing (ZF)-based element-wise sequential optimization (EWSO) method is employed to obtain a favorable initial point \cite{pa8}. Specifically, each waveguide segment is uniformly discretized into $N_{\rm ini}$ candidate positions, where $N_{\rm ini}$ is kept small to limit the initialization complexity. The transmit and receive PA positions are then sequentially optimized by searching these candidates while fixing the remaining PA positions. For each candidate, the beamforming matrix is constructed using the ZF-based method detailed in Appendix~C, and the corresponding CRLB is evaluated. Candidates violating the PA spacing or communication rate constraints are discarded, and the feasible candidate yielding the lowest CRLB is retained. After a single sweep over all PAs, the resulting PA positions and beamforming matrix are used to initialize the RMO method for further joint optimization. Based on the above, the overall RMO method for solving \eqref{Q2} is summarized in \textbf{Algorithm 2}.
	
	For fixed penalty and smoothing parameters, the Armijo line-search strategy \eqref{Amijo} ensures sufficient descent of the smooth penalized objective along a descent direction. The outer iterations progressively adjust these parameters to improve constraint satisfaction. Since the resulting problem is non-convex, the proposed method is intended to obtain a numerically feasible solution rather than a globally optimal solution.
	
	\subsection{Computational Complexity Analysis}
	The computational cost of \textbf{Algorithm 2} mainly arises from the evaluations of the CRLB and communication rates and their Euclidean gradients, as well as the RBFGS updates. For the CRLB calculation, constructing the sensing channels requires $\mathcal{O}(NMK_T)$ operations. Using the Jacobian $\mathbf{Z}\in\mathbb{C}^{M(K_C+1)\times 4K_T}$, forming the FIM and its adjoint requires $\mathcal{O}(M(K_C+1)K_T^2)$ operations, while evaluating the Schur complement and its inverse requires $\mathcal{O}(K_T^3)$ operations. The remaining gradient computations require $\mathcal{O}(M(K_C+1)K_T+NMK_T)$ operations. Thus, jointly evaluating the CRLB and its gradients has complexity $\mathcal{O}(NMK_T+M(K_C+1)K_T^2+K_T^3)$. Computing all CU rates and their gradients requires $\mathcal{O}(NMK_C+MK_C(K_C+1))$ operations. The minimum distance penalties of the TPAs and their gradients require $\mathcal{O}(MN^2)$ operations, while \textbf{Algorithm 1} requires $\mathcal{O}(S(M(K_C+1)+NM))$ operations per iteration. The ZF-based EWSO initialization examines $\mathcal{O}(MNN_{\rm ini})$ PA candidates, each requiring $\mathcal{O}(C_{\rm ZF})$ operations for ZF construction, null-space sensing beam design, and CRLB evaluation of at most $K_C+2$ power allocations, where
	\begin{equation}
		C_{\rm ZF}\!\!=\!\!\mathcal{O}\left(
		M^3K_T\!+\!NMK_CK_T\!+\!MK_C^2K_T^2\!+\!K_CK_T^3
		\right).
	\end{equation}
	Then, the overall complexity of \textbf{Algorithm 2} is estimated as
	\begin{align}
		\mathcal{O}\Bigl(&MNN_{\rm ini}C_{\rm ZF}
		+L_{\max}I_{\max}\bigl[
		K_T^3+M\bigl(
		N^2+K_C^2 \nonumber\\
		&+K_CK_T^2+N(K_T+K_C)
		+S(N+K_C)
		\bigr)\bigr]\Bigr).
	\end{align}
	Since $N_{\rm ini}$ is set to a small constant, it does not affect the asymptotic complexity order and introduces only limited initialization overhead.

	\section{Simulation Results}
	In this section, we verify the performance of the proposed SWAN-assisted ISAC systems and the effectiveness of the proposed RMO method. Unless otherwise specified, the simulation parameters are set as follows. The side lengths of the service area are $D_x=80$ and $D_y=40$, respectively. The height of the SWAN is $d=3$. The carrier frequency is $f_c = \frac{3\times 10^8}{\lambda} = 28$ GHz, the effective refractive index of the waveguide is $\eta_e = 1.4$, the average attenuation factor along the waveguide is $\kappa = 0.08$ dB/m. The noise powers for sensing at the BS and for communications at the CUs are set as $\sigma_s^2 = \sigma_c^2= -90$ dBm. The transmit power is $P_t=26$ dBm. The number of TSWs and RSWs is $M=10$, and the number of TPAs activated on each TSW is $N=4$. The number of CUs and targets is $K_C=K_T=4$. The number of signal symbols per ISAC period is $T=512$. The $x$ and $y$ coordinates of the CUs and targets are randomly generated within the ranges of $[0,D_x]$ and $[-\frac{D_y}{2},\frac{D_y}{2}]$, respectively. The target scattering coefficients are generated as $\alpha_k=e^{j2\pi\mathcal{U}(0,1)}$. The rate constraint for all the CUs is $\Gamma_k = \Gamma = 4\ \text{bps/Hz}, \forall k\in\mathcal{K}_C$. For the RMO method,  $N_{\rm ini} = 10$ is set for initialization, the update factors are set as $\rho^0=1$ and $\theta_\rho=3$, and the initial smoothing parameter and corresponding update weight and minimum value are $u^0=0.1$, $\theta_u=0.5$, and $u_{\min}=10^{-6}$. The scaling factor of the sigmoid function is set as $\mu = 10^{-3}$. The convergence threshold is $\tau=10^{-6}$. The size of the limited memory is $S=30$.

	The proposed scheme is referred to as “\textbf{SWAN-RMO}”. To evaluate the effectiveness of the proposed SWAN-assisted ISAC system and the corresponding RMO method under different system parameter settings, we consider the following baseline schemes for performance comparisons, and all the results are averaged over 1024 random realizations of CU and target positions.
	\begin{itemize}
		\item ``\textbf{MIMO-RMO}": The conventional MIMO architecture is adopted, where $M$ RF chains are connected to the position of the feed points of the TSW. Each RF chain is connected to $N$ antennas arranged with half-wavelength spacing for signal transmission, while $M$ receive antennas are placed at the RSW feed points. The beamforming $\mathbf{W}$ is obtained by the proposed RMO with fixed $\boldsymbol{\phi}$ and $\boldsymbol{\psi}$.
		\item ``\textbf{MPASS-RMO}": Existing MPASS is adopted for the ISAC \cite{pa4isac2,pa4isac3}, where $M$ waveguides are used for transmission and $M$ waveguides are used for reception. All waveguide feed points are located at $x=0$, with the $y$-coordinates of their feed points uniformly distributed within $[-20,20]$. The beamforming and PA positions are obtained via the proposed RMO method. Compared with the SWAN scheme, the MPASS scheme incurs an additional waveguide deployment length of $2(M-1)D_x$ meters.
		\item ``\textbf{SWAN-RMO with MP}": For the SWAN scheme, the TPA and RPA positions are determined using midpoint-based placement to reduce propagation distances \cite{pa4isac2, swan1}. Specifically, the reference TPA position is obtained from the centroid of the CU and target locations and projected onto each corresponding waveguide segment, while the reference RPA position is determined from the centroid of the target locations. The TPAs are placed around their reference positions subject to the minimum spacing and segment boundary constraints. The beamforming matrix is then optimized using the proposed RMO method with fixed PA positions.
		\item ``\textbf{SWAN-ZF-based EWSO}": For the SWAN scheme, the ZF-based EWSO method introduced in Section IV-D is adopted as a baseline without subsequent RMO optimization. A larger number of candidate positions $N_{\rm ini}$ is adopted for each waveguide segment to enable a finer search. The search resolution is selected to achieve a runtime comparable to that of the proposed method for fair comparison.
		
	\end{itemize}
	\begin{figure}[t]
		\centering
		{\includegraphics[width=0.9\columnwidth]{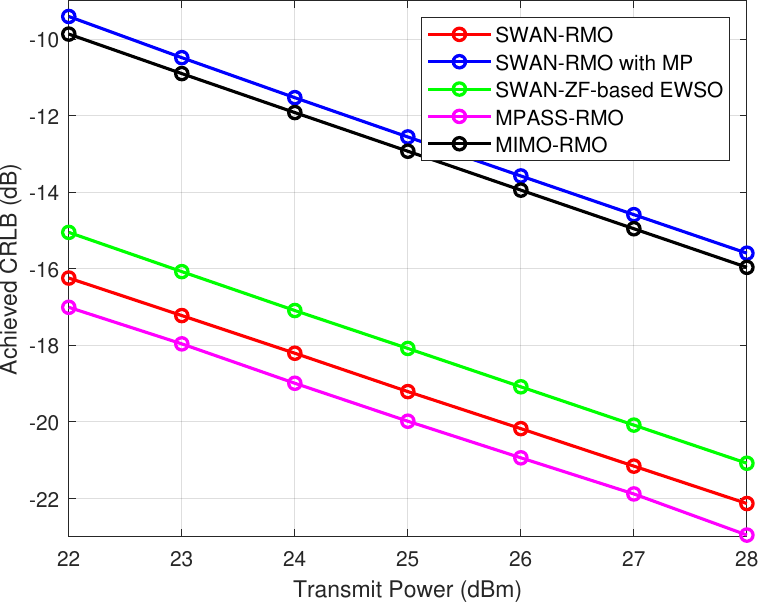}}
		\vspace{-0.3 cm}
		\caption{The achieved CRLB versus the transmit power $P_t$.}
		\vspace{-0.3 cm}
	\end{figure}
	
	\subsubsection{Impact of the Transmit Power}
	Figure 2 shows the performance of different schemes versus the transmit power. It is observed that the CRLBs of all schemes decrease as the transmit power increases, since a larger power budget allows more power to be allocated to sensing while satisfying the communication rate constraints. Notably, the “SWAN-RMO with MP” scheme, which optimizes the PA positions by reducing the average propagation distance as commonly considered in single-target or single-CU scenarios, performs even worse than the “MIMO-RMO” scheme with fixed antenna positions, despite employing the proposed RMO method for beamforming optimization. This is because reducing the average path loss alone does not adequately account for inter-user interference in multi-target and multi-user scenarios. Moreover, the additional in-waveguide propagation loss may offset the benefits of flexible PA positioning, resulting in degraded CRLB performance. The proposed ``SWAN-RMO" scheme achieves approximately a 6 dB reduction in CRLB compared with the conventional MIMO scheme. This improvement is attributed to the joint optimization of beamforming and PA positions, which better exploits the additional spatial degrees of freedom provided by flexible PA deployment to mitigate inter-user interference and enhance sensing performance. Furthermore, compared with the ``ZF-based EWSO" scheme, the proposed RMO method achieves an additional CRLB reduction of approximately 1.1 dB. This is because the simultaneous and joint optimization of beamforming and PA positions better exploits their coupling, enabling a more effective trade-off between communication rate requirements and sensing performance.
	
	It is further observed that the ``MPASS-RMO" scheme achieves a slightly lower CRLB than the proposed ``SWAN-RMO" scheme, with a performance gain of approximately 0.75 dB. This is mainly because the deployment of multiple full-length waveguides in MPASS provides more favorable propagation conditions for both targets and CUs, thereby reducing large-scale path loss. However, MPASS requires an additional total waveguide length of $2(M-1)D_x$ compared with SWAN, resulting in higher hardware deployment and maintenance costs. Therefore, the proposed SWAN architecture offers a more deployment-efficient alternative by substantially reducing the required waveguide length at the expense of only a marginal degradation in CRLB performance.
	
	\subsubsection{Impact of the Length of the Service Area}
	\begin{figure}[t]
		\centering
		{\includegraphics[width=0.9\columnwidth]{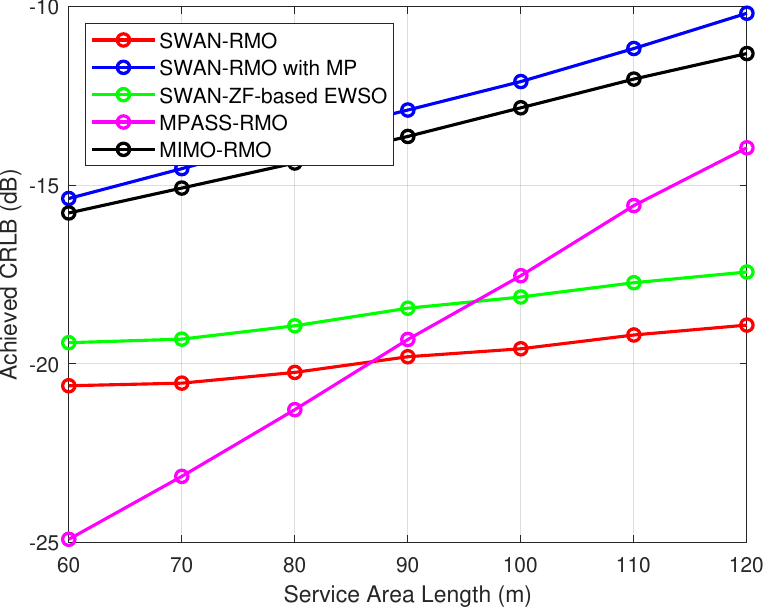}}
		\vspace{-0.3 cm}
		\caption{The achieved CRLB versus the service area length $D_x$.}
		\vspace{-0.3 cm}
	\end{figure}
	The size of the service area also affects the CRLB performance. Figure 3 illustrates the CRLB achieved by different schemes versus the service area length $D_x$. It is observed that the CRLBs of all schemes increase with $D_x$, mainly due to increased free-space path loss and additional in-waveguide propagation loss in waveguide-enabled schemes. Consistent with the previous results, the proposed joint beamforming and PA position optimization scheme outperforms both the conventional MIMO scheme and the scheme employing midpoint-based PA positioning with respect to the CUs and targets.
	
	More importantly, the relative performance of ``MPASS-RMO" and ``SWAN-RMO" varies with $D_x$. When $D_x$ is small, ``MPASS-RMO" achieves a more noticeable CRLB improvement over ``SWAN-RMO", since its multiple full-length waveguides provide favorable propagation conditions with relatively low free-space and in-waveguide path losses. However, as $D_x$ increases, the CRLB of “MPASS-RMO” increases more rapidly and eventually exceeds that of “SWAN-RMO”. This is mainly attributed to the increased in-waveguide propagation loss along the longer waveguides in MPASS, particularly for PAs located far from their feed points. In contrast, the segmented architecture of SWAN maintains relatively short propagation distances between the feed points and PAs, thereby mitigating the in-waveguide loss and making its CRLB performance less sensitive to increasing $D_x$. These results demonstrate that SWAN offers a more favorable sensing performance than MPASS for larger service areas, where longer waveguide deployments are required.

	\subsubsection{Impact of the CU Number and Communication Rate Threshold}
	\begin{figure}[t]
		\centering
		{\includegraphics[width=0.94\columnwidth]{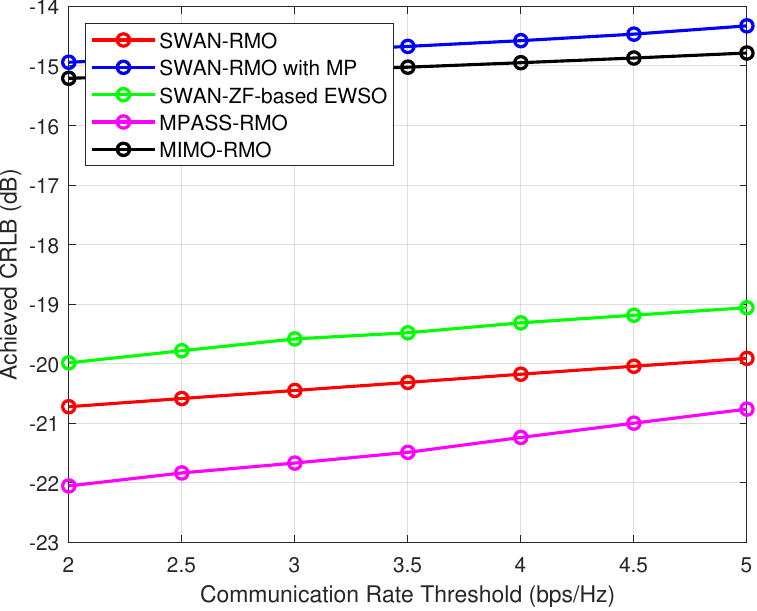}}
		\vspace{-0.3 cm}
		\caption{The achieved CRLB versus the communication rate threshold $\Gamma$.}
		\vspace{-0.3 cm}
	\end{figure}
	\begin{figure}[t]
		\centering
		{\includegraphics[width=0.9\columnwidth]{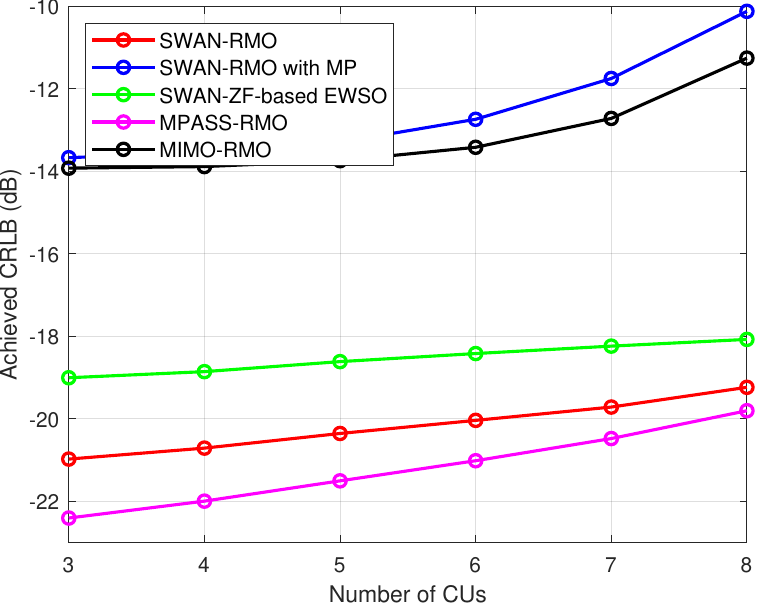}}
		\vspace{-0.3 cm}
		\caption{The achieved CRLB versus the number of CUs $K_C$.}
		\vspace{-0.3 cm}
	\end{figure}
	Figure 4 illustrates the achieved CRLB versus the communication rate threshold $\Gamma$. It is observed that the CRLBs of all schemes increase with $\Gamma$. This is because, under the adopted dual-functional signal design, a higher communication rate requirement generally necessitates allocating more power resources to satisfy the communication constraints, thereby limiting the power available for sensing. Consistent with the previous observations, the proposed SWAN-RMO scheme continues to outperform the other SWAN-based schemes, demonstrating the effectiveness of jointly optimizing beamforming and PA positions. Moreover, SWAN-RMO achieves a slightly higher CRLB than MPASS-RMO, but requires substantially less waveguide deployment, highlighting its favorable trade-off between sensing performance and hardware overhead.
	
	Figure 5 illustrates the achieved CRLB versus the number of CUs. It is observed that the CRLBs of all schemes increase with the number of CUs, since satisfying the communication rate requirements of more users generally requires additional power resources, thereby limiting the power available for sensing. Notably, the CRLBs of the conventional MIMO scheme and the scheme employing CU- and target-based PA positioning increase more rapidly as the number of CUs grows. In contrast, the proposed SWAN-RMO scheme exhibits a slower increase in CRLB, demonstrating that the joint optimization of beamforming and PA positions can better exploit the available spatial degrees of freedom to manage inter-user interference and balance communication and sensing requirements in multi-user scenarios.
	
	\subsubsection{Impact of Target Location Errors}
	\begin{figure}[t]
		\centering
		{\includegraphics[width=0.9\columnwidth]{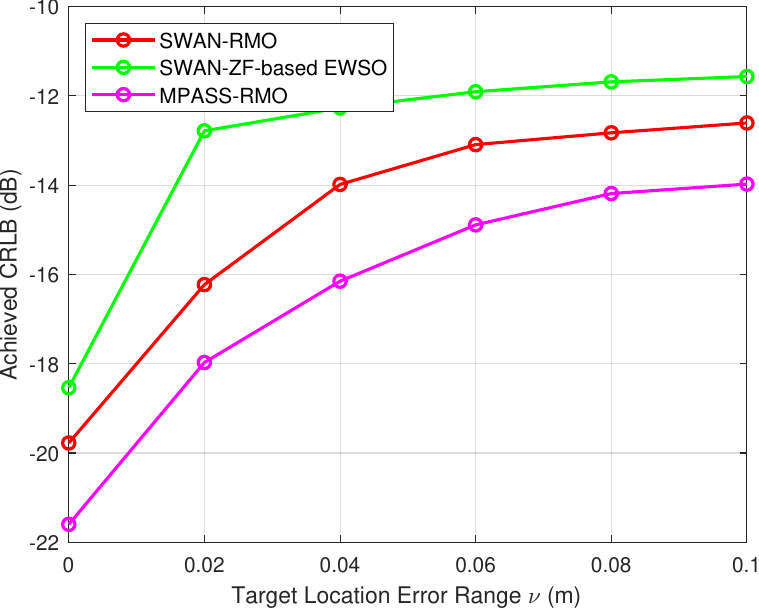}}
		\vspace{-0.3 cm}
		\caption{The achieved CRLB versus the target location error range $\nu$.}
		\vspace{-0.3 cm}
	\end{figure}
	As discussed in Section III-A, the target location information $\boldsymbol{x}$ and $\boldsymbol{y}$ may be imperfect for the CRLB minimization due to practical concerns such as target movement or estimation error in the previous sensing stage. We test the effectiveness of the proposed scheme under the imperfect target information condition in this subsection. We assume that the real locations of the targets are given as $\tilde{x}_{t,k}\sim\mathcal{U}(x_{t,k}-\nu/2,x_{t,k}+\nu/2)$ and $\tilde{y}_{t,k}\sim\mathcal{U}(y_{t,k}-\nu/2,y_{t,k}+\nu/2)$, where $\nu$ is the location error. A larger $\nu$ corresponds to a larger target location information error. Figure 6 illustrates the achieved CRLB versus the target location error range $\nu$. It is observed that the ``SWAN-ZF-based EWSO" scheme that employs sequential PA position optimization with ZF-based beamforming suffers a more pronounced CRLB degradation as $\nu$ increases. In contrast, the proposed SWAN-RMO scheme exhibits a slower increase in CRLB under relatively small target position errors, demonstrating the benefits of jointly optimizing beamforming and PA positions. Moreover, when the target position error becomes relatively large, the MPASS scheme exhibits better robustness than the SWAN scheme. This can be attributed to the more distributed waveguide deployment in MPASS, which provides favorable propagation conditions for the targets and helps mitigate the impact of position errors on sensing performance. Nevertheless, this improved robustness comes at the expense of increased hardware deployment and maintenance costs due to the substantially longer total waveguide length required by MPASS.

	\vspace{-0.2 cm}
	\section{Conclusion}
	In this paper, we have investigated a novel SWAN-assisted multi-user multi-target ISAC system. We adopted the CRLB for target location estimation as the sensing metric and proposed an RMO method to solve the rate-constrained CRLB minimization problem by jointly optimizing the beamforming and PA positions. Simulation results have verified the effectiveness of the proposed RMO method and the advantages of the SWAN scheme for the multi-user multi-target ISAC systems compared with baseline schemes, including conventional MIMO, MPASS, and separate optimization schemes. It should be noted that this work mainly focuses on the theoretical performance of the SWAN-assisted ISAC systems. Practical issues, such as more effective algorithms for real-time and rapid control of the beamforming and PA positions, could be focused on for future work.

	\appendices
	\section{Expressions of the FIM and Equivalent FIM}
	
	We first obtain the derivatives of $\boldsymbol{\Psi}$ with respect to the target location parameters as
	\begin{align}
		\frac{\partial\boldsymbol{\Psi}}
		{\partial x_{t,k}}
		&=\operatorname{vec}
		\left(\dot{\mathbf H}_{x_{t,k}}\mathbf X\right),
		\\
		\frac{\partial\boldsymbol{\Psi}}
		{\partial y_{t,k}}
		&=\operatorname{vec}
		\left(\dot{\mathbf H}_{y_{t,k}}\mathbf X\right),
	\end{align}
	where
	\begin{align}
		\dot{\mathbf H}_{x_{t,k}}
		=\frac{\partial\mathbf H_{t,k}}
		{\partial x_{t,k}}=\alpha_k
		\frac{\partial\mathbf h_{r,k}}
		{\partial x_{t,k}}\mathbf h_{t,k}^{\top}
		+\alpha_k\mathbf h_{r,k}
		\frac{\partial\mathbf h_{t,k}^{\top}}
		{\partial x_{t,k}}
	\end{align}
	and
	\begin{align}
		\dot{\mathbf H}_{y_{t,k}}
		=\frac{\partial\mathbf H_{t,k}}
		{\partial y_{t,k}}=\alpha_k
		\frac{\partial\mathbf h_{r,k}}
		{\partial y_{t,k}}\mathbf h_{t,k}^{\top}
		+\alpha_k\mathbf h_{r,k}
		\frac{\partial\mathbf h_{t,k}^{\top}}
		{\partial y_{t,k}}.
	\end{align}
	The channel derivatives are given by
	\begin{align}
		\frac{\partial\mathbf h_{r,k}[m]}
		{\partial x_{t,k}}
		\!=\!-a_m(x_{t,k}-\phi_m)b_{m,k}
		\bigg(
		{D_{t,\phi}^{k,m}}^{-3}+\jmath k_c
		{D_{t,\phi}^{k,m}}^{-2}
		\bigg),\nonumber
	\end{align}
	\begin{align}
		\frac{\partial\mathbf h_{r,k}[m]}
		{\partial y_{t,k}}
		\!=\!-a_m(y_{t,k}-y_{\rm rsw})b_{m,k}
		\bigg(
		{D_{t,\phi}^{k,m}}^{-3}\!+\!\jmath k_c
		{D_{t,\phi}^{k,m}}^{-2}
		\bigg),\nonumber
	\end{align}
	\begin{align}
		\frac{\partial\mathbf h_{t,k}^{\top}[m]}
		{\partial x_{t,k}}
		\!\!=\!\!-\!\!\sum_{n=1}^{N}
		\!c_m^n(x_{t,k}\!\!-\!\!\psi_m^n)d_{m,k}^n\!\bigg(\!\!
		{D_{t,\psi}^{k,m,n}}^{-3}
		\!\!\!+\!\!\jmath k_c
		{D_{t,\psi}^{k,m,n}}^{-2}
		\!\bigg),\nonumber
	\end{align}
	and
	\begin{align}
		\frac{\partial\mathbf h_{t,k}^{\top}[m]}
		{\partial y_{t,k}}
		\!\!=\!\!-\!\!\!\sum_{n=1}^{N}\!
		c_m^n(y_{t,k}\!\!-\!\!y_{\rm tsw})d_{m,k}^n\!\!\bigg(\!\!
		{D_{t,\psi}^{k,m,n}}^{-3}\!\!
		+\!\!\jmath k_c
		{D_{t,\psi}^{k,m,n}}^{-2}
		\!\bigg),\nonumber
	\end{align}
	where $a_m=\eta^{\frac12}10^{-\frac{\kappa}{20}(\phi_m-\phi_m^0)}$, $c_m^n=\sqrt{\frac{\eta}{N}}10^{-\frac{\kappa}{20}(\psi_m^n-\psi_m^0)}$, $b_{m,k}=e^{-\jmath\left(k_cD_{t,\phi}^{k,m}+\frac{2\pi}{\lambda_g}(\phi_m-\phi_m^0)\right)}$, and $d_{m,k}^n=e^{-\jmath\left(k_cD_{t,\psi}^{k,m,n}+\frac{2\pi}{\lambda_g}(\psi_m^n-\psi_m^0)\right)}$. To derive the derivatives with respect to the unknown target scattering coefficients, we define $\mathbf A_k=\mathbf h_{r,k}\mathbf h_{t,k}^{\top}$ such that $\mathbf H_{t,k}=\alpha_k\mathbf A_k$, then we have
	\begin{align}
		\frac{\partial\boldsymbol{\Psi}}
		{\partial \boldsymbol{\alpha}_{R}[k]}
		=\operatorname{vec}(\mathbf A_k\mathbf X),
		\frac{\partial\boldsymbol{\Psi}}
		{\partial\boldsymbol{\alpha}_{I}[k]}
		=\jmath\operatorname{vec}(\mathbf A_k\mathbf X).
	\end{align}
	
	Based on the above derivatives and the FIM definition, the location-related FIM submatrices are obtained as
	\begin{align}
		\mathbf F_{\boldsymbol{x}\boldsymbol{x}^{\top}}
		[i,j]
		=
		\frac{2T}{\sigma_s^2}
		\Re\left\{
		\operatorname{Tr}\left(
		\dot{\mathbf H}_{x_{t,i}}
		\mathbf W\mathbf W^{\rm H}
		\dot{\mathbf H}_{x_{t,j}}^{\rm H}
		\right)\right\},
	\end{align}
	\begin{align}
		\mathbf F_{\boldsymbol{x}\boldsymbol{y}^{\top}}
		[i,j]
		=
		\frac{2T}{\sigma_s^2}
		\Re\left\{
		\operatorname{Tr}\left(
		\dot{\mathbf H}_{x_{t,i}}
		\mathbf W\mathbf W^{\rm H}
		\dot{\mathbf H}_{y_{t,j}}^{\rm H}
		\right)\right\},
	\end{align}
	and
	\begin{align}
		\mathbf F_{\boldsymbol{y}\boldsymbol{y}^{\top}}
		[i,j]=
		\frac{2T}{\sigma_s^2}
		\Re\left\{
		\operatorname{Tr}\left(
		\dot{\mathbf H}_{y_{t,i}}
		\mathbf W\mathbf W^{\rm H}
		\dot{\mathbf H}_{y_{t,j}}^{\rm H}
		\right)\right\}.
	\end{align}
	We then define
	\begin{align}
		\mathbf E_i=
		\begin{cases}
			\dot{\mathbf H}_{x_{t,i}},
			&1\leq i\leq K_T,\\
			\dot{\mathbf H}_{y_{t,i-K_T}},
			&K_T<i\leq 2K_T
		\end{cases}
	\end{align}
	and introduce matrices
	$\mathbf U\in\mathbb C^{2K_T\times2K_T}$,
	$\mathbf R\in\mathbb C^{2K_T\times K_T}$,
	and $\mathbf G\in\mathbb C^{K_T\times K_T}$,
	whose elements are respectively given by $\mathbf U[i,j]=\operatorname{Tr}\left(\mathbf E_i^{\rm H}\mathbf E_j\mathbf W\mathbf W^{\rm H}\right)$, $\mathbf R[i,j]=\operatorname{Tr}\left(\mathbf E_i^{\rm H}\mathbf A_j\mathbf W\mathbf W^{\rm H}\right)$, $\mathbf G[i,j]=\operatorname{Tr}\left(\mathbf A_i^{\rm H}\mathbf A_j\mathbf W\mathbf W^{\rm H}\right)$.	Then we obtain the FIMs as
	\begin{align}
		\mathbf F_{\xi\xi}
		&=\frac{2T}{\sigma_s^2}
		\Re\{\mathbf U\},\\
		\mathbf F_{\xi\alpha}
		&=\frac{2T}{\sigma_s^2}
		\begin{bmatrix}
			\Re\{\mathbf R\}&-\Im\{\mathbf R\}
		\end{bmatrix},\\
		\mathbf F_{\alpha\alpha}
		&=\frac{2T}{\sigma_s^2}
		\begin{bmatrix}
			\Re\{\mathbf G\}&-\Im\{\mathbf G\}\\
			\Im\{\mathbf G\}&\Re\{\mathbf G\}
		\end{bmatrix},
	\end{align}
	where
	$\mathbf F_{\alpha\xi}
	=\mathbf F_{\xi\alpha}^{\top}$. By applying the Schur complement and the real-valued representation of complex matrices, the equivalent FIM for target localization is given by
	\begin{align}
		\mathbf F_{\rm eq}
		\!=\!\mathbf F_{\xi\xi}
		\!-\!\mathbf F_{\xi\alpha}
		\mathbf F_{\alpha\alpha}^{-1}
		\mathbf F_{\alpha\xi} =\frac{2T}{\sigma_s^2}
		\Re\left\{
		\mathbf U\!-\!
		\mathbf R\mathbf G^{-1}\mathbf R^{\rm H}
		\right\}.
	\end{align}
	
	\section{Derivations of Closed-Form Euclidean Gradients}
	The Euclidean gradient derivations with optimized computational complexity are given as follows.
	\subsection{Euclidean Gradients of the CRLB Objective}
	Let $f(\mathbf{X})=f=\operatorname{CRLB}_{\boldsymbol{\xi}}$ and
	$\beta_s=2T/\sigma_s^2$. Following Appendix A, define
	$\mathbf A_k=\mathbf h_{r,k}\mathbf h_{t,k}^{\top}$ and
	\begin{align}
		\mathbf z_{q,k}&=\operatorname{vec}(\dot{\mathbf H}_{q_{t,k}}\mathbf W),
		\quad q\in\{x,y\},\\
		\mathbf z_{a,k}&=\operatorname{vec}(\mathbf A_k\mathbf W),
	\end{align}
	where $\mathbf W=\sqrt{P_t}\tilde{\mathbf W}$. Collecting these
	vectors in the parameter order
	$[\boldsymbol{x}^{\top},\boldsymbol{y}^{\top},
	\boldsymbol{\alpha}_R^{\top},\boldsymbol{\alpha}_I^{\top}]^{\top}$ gives
	\begin{align}
		\mathbf Z=\big[&\mathbf z_{x,1},\ldots,\mathbf z_{x,K_T},
		\mathbf z_{y,1},\ldots,\mathbf z_{y,K_T},\nonumber\\
		&\mathbf z_{a,1},\ldots,\mathbf z_{a,K_T},
		\jmath\mathbf z_{a,1},\ldots,\jmath\mathbf z_{a,K_T}\big],
	\end{align}
	so that $\mathbf F_{\theta}=\beta_s\Re\{\mathbf Z^{\mathrm H}\mathbf Z\}$.
	Using the FIM partition in Section III, set $\mathbf J=\mathbf F_{\alpha\alpha}^{-1}\mathbf F_{\alpha\xi}$, $\mathbf T_0=\mathbf F_{\rm eq}^{-1}$, $\mathbf V = [\mathbf I_{2K_T},-\mathbf J]^\top\mathbf T_0$, and $\mathbf Q=-\mathbf Z\mathbf V\mathbf V^{\top}$, $f=\operatorname{Tr}(\mathbf T_0)$ and its differential is
	\begin{equation}
		\mathrm df=2\beta_s\Re\{\operatorname{Tr}(\mathbf Q^{\mathrm H}
		\mathrm d\mathbf Z)\}.
	\end{equation}
	Let $\mathbf Q_{x,k}$, $\mathbf Q_{y,k}$, $\mathbf Q_{a,k}$,
	and $\mathbf Q_{b,k}$ be the $M\times (K_C+1)$ matrices obtained by
	reshaping the corresponding columns of $\mathbf Q$, and set
	$\mathbf Q_{c,k}=\mathbf Q_{a,k}-\jmath\mathbf Q_{b,k}$.
	The Euclidean beamforming gradient is
	\begin{align}
		\nabla_{\tilde{\mathbf W}^{*}}f
		=2\beta_s\sqrt{P_t}\sum_{k=1}^{K_T}\big(
		\dot{\mathbf H}_{x_{t,k}}^{\mathrm H}\mathbf Q_{x,k}
		+\dot{\mathbf H}_{y_{t,k}}^{\mathrm H}\mathbf Q_{y,k}
		+\mathbf A_k^{\mathrm H}\mathbf Q_{c,k}\big).
	\end{align}
	
	For the PA position gradients, let
	$\mathbf u_k=\mathbf h_{r,k}$,
	$\mathbf v_k=\mathbf h_{t,k}$,
	and $\mathbf u_{q,k}=\partial\mathbf h_{r,k}/\partial q_{t,k}$,
	$\mathbf v_{q,k}=\partial\mathbf h_{t,k}/\partial q_{t,k}$,
	$q\in\{x,y\}$. Define $\mathbf e_{0,k}=\sum_{q\in\{x,y\}}
	\mathbf u_{q,k}^{\mathrm H}\mathbf Q_{q,k}$, $\mathbf e_{q,k}=\mathbf u_k^{\mathrm H}\mathbf Q_{q,k},
	q\in\{x,y,c\}$, $\mathbf g_{0,k}=\mathbf v_k^{\top}\mathbf W$, and $\mathbf g_{q,k}=\mathbf v_{q,k}^{\top}\mathbf W,
	\quad q\in\{x,y\}$, the transmit- and receive-side coefficient vectors are
	\begin{align}
		\mathbf b_{0,k}^{t}
		&=\alpha_k(\mathbf e_{0,k}\mathbf W^{\mathrm H})^{*}
		+(\mathbf e_{c,k}\mathbf W^{\mathrm H})^{*},\\
		\mathbf b_{q,k}^{t}
		&=\alpha_k(\mathbf e_{q,k}\mathbf W^{\mathrm H})^{*},
		\quad q\in\{x,y\},\\
		\mathbf b_{0,k}^{r}
		&=\alpha_k\bigg(\sum_{q\in\{x,y\}}
		\mathbf Q_{q,k}\mathbf g_{q,k}^{\mathrm H}\bigg)^{*}
		+(\mathbf Q_{c,k}\mathbf g_{0,k}^{\mathrm H})^{*},\\
		\mathbf b_{q,k}^{r}
		&=\alpha_k(\mathbf Q_{q,k}\mathbf g_{0,k}^{\mathrm H})^{*},
		\quad q\in\{x,y\}.
	\end{align}
	For a transmit or receive PA at $a\in\{\psi_m^n,\phi_m\}$,
	let $r$ denote its distance to target $k$,
	$\Delta_x=x_{t,k}-a$, and
	$\Delta_y=y_{t,k}-y_{\rm PA}$, where
	$y_{\rm PA}=y_{\rm tsw}$ or $y_{\rm rsw}$, respectively.
	Under $a\ge a_0$, define
	\begin{align}
		\Gamma&=-\frac{\kappa\ln 10}{20}
		+\frac{\Delta_x}{r^2}
		+\jmath k_c\frac{\Delta_x}{r}
		-\jmath\frac{2\pi}{\lambda_g},\\
		\chi_q&=-\Delta_q\left(\frac{1}{r^2}
		+\frac{\jmath k_c}{r}\right),\quad q\in\{x,y\},\\
		\chi'_x&=\frac{1}{r^2}-\frac{2\Delta_x^2}{r^4}
		+\jmath k_c\left(\frac{1}{r}-\frac{\Delta_x^2}{r^3}\right),\\
		\chi'_y&=-\Delta_x\Delta_y\left(\frac{2}{r^4}
		+\frac{\jmath k_c}{r^3}\right).
	\end{align}
	Thus, the elemental channel coefficient $h$ obeys
	$\partial h/\partial a=h\Gamma$ and
	$\partial(h\chi_q)/\partial a=h(\Gamma\chi_q+\chi'_q)$.
	For $s\in\{t,r\}$, let $h_k^s(a)$ be the corresponding
	TPA--target or target--RPA channel coefficient, and define
	\begin{align}
		\mathcal D_{k,m}^{s}(a)=h_k^s(a)\bigg[&
		\Gamma b_{0,k,m}^{s}
		+\sum_{q\in\{x,y\}}
		(\Gamma\chi_q+\chi'_q)b_{q,k,m}^{s}\bigg],
	\end{align}
	where $b_{q,k,m}^{s}=[\mathbf b_{q,k}^{s}]_m$,
	$h_k^t(\psi_m^n)=c_m^n d_{m,k}^n/D_{t,\psi}^{k,m,n}$, and
	$h_k^r(\phi_m)=a_m b_{m,k}/D_{t,\phi}^{k,m}$, as defined in Appendix A.
	The Euclidean position gradients are
	\begin{align}
		\nabla_{\psi_m^n}f
		&=2\beta_s\Re\bigg\{\sum_{k=1}^{K_T}
		\mathcal D_{k,m}^{t}(\psi_m^n)\bigg\},\\
		\nabla_{\phi_m}f
		&=2\beta_s\Re\bigg\{\sum_{k=1}^{K_T}
		\mathcal D_{k,m}^{r}(\phi_m)\bigg\}.
	\end{align}
	Finally, the element-wise sigmoid projections in
	\eqref{projpsi} and \eqref{projphi} yield
	\begin{align}
		\nabla_{\tilde{\boldsymbol a}} f
		=L\mu\,\nabla_{\boldsymbol a} f
		\odot\operatorname{sig}(\tilde{\boldsymbol a})
		\odot\big[\mathbf 1-\operatorname{sig}(\tilde{\boldsymbol a})\big],
		\quad \boldsymbol a\in\{\boldsymbol\psi,\boldsymbol\phi\}.
	\end{align}
	
	\subsection{Euclidean Gradients of Communication Rates Constraints}
	For $\boldsymbol{\upsilon} \in \{\tilde{\mathbf{W}}, \tilde{\boldsymbol{\psi}}, \tilde{\boldsymbol{\phi}}\}$, we have $\nabla_{\boldsymbol{\upsilon}}h_i(\mathbf{X}) = -\nabla_{\boldsymbol{\upsilon}}R_i, \forall i\in\mathcal{K}_C$. Define $u_{k,j} = \mathbf{h}_{c,k}^\top \mathbf{w}_j$, $j\in\mathcal{K}_C\cup\{S\}$, signal power $S_k = |u_{k,k}|^2$, interference-plus-noise $I_k = \sum_{j\in\mathcal{K}_C\setminus\{k\}} |u_{k,j}|^2+|u_{k,S}|^2 + \sigma^2_c$, and $\text{SINR}_k = S_k / I_k$. The gradients are computed through the following sequence. First, we calculate the rate chain-rule coefficient matrix $\mathbf{C} \in \mathbb{C}^{K_C \times (K_C+1)}$ as
	\begin{equation}
		C_{k,j} = \frac{1}{\ln(2) (1 + \gamma_k)} \times
		\begin{cases} 
			\frac{2 u_{k,k}}{I_k}, & j = k \\ 
			-\frac{2 u_{k,j} S_k}{I_k^2}, & j \neq k 
		\end{cases},
	\end{equation}
	then the gradient with respect to $\mathbf{w}_j$ is exactly given by
	\begin{equation}
		\nabla_{\mathbf{w}^*_j} R_k = \mathbf{h}_{c,k}^* C_{k,j}, 
	\end{equation}
	then we can obtain $\nabla_{{\tilde{\mathbf{W}}}} R_k = \sqrt{P_t}\nabla_{\mathbf{W}} R_k$.
	
	For the gradient with respect to the TPA positions, let $\Delta x_c^{n,m,k} = x_{c,k} - \psi_m^n$. Compute the geometric-phase multiplier as
	\begin{equation}
		G_{n,m,k} = -\frac{\kappa \ln 10}{20} + \frac{\Delta x_c^{n,m,k}}{(D_{c,\psi}^{k,m,n})^2} + \jmath k_c \frac{\Delta x_c^{n,m,k}}{D_{c,\psi}^{k,m,n}} - \jmath \frac{2\pi}{\lambda_g}
	\end{equation}
	The Euclidean gradient mapped to the position variable $\psi_m^n$ is then obtained as
	\begin{equation}
		\nabla_{\psi_m^n} R_k = \operatorname{Re} \left\{ \left( h_{n,m,k} \cdot G_{n,m,k} \right)\!\!\!\!\!\! \sum_{j\in\mathcal{K}_C\cup\{S\}}\!\!\!\!\!\! \mathbf{w}_j[m] C^*_{k,j} \right\},
	\end{equation}
	where
	\begin{equation}
		h_{n,m,k} = \frac{\sqrt{\eta}10^{-\frac{\kappa}{20}(\psi_m^n-\psi_m^0)}}{\sqrt{N}D_{c,\psi}^{k,m,n}}  e^{-j\left(k_c D_{c,\psi}^{k,m,n} + \frac{2\pi}{\lambda_g}(\psi_m^n-\psi_m^0)\right)},
	\end{equation}
	then we obtain the gradient
	\begin{align}
		\nabla_{\tilde{\boldsymbol{\psi}}} R_k =& \nabla_{{\boldsymbol{\psi}}} R_k\odot L\mu\left[\operatorname{sig}(\tilde{\psi}_1^1),\dots,\operatorname{sig}(\tilde{\psi}_M^N)\right]\nonumber\\&\quad\odot \left[(1-\operatorname{sig}(\tilde{\psi}_1^1)),\dots,(1-\operatorname{sig}(\tilde{\psi}_M^N))\right].
	\end{align}

	\section{ZF-Based Beamforming for Initialization}
	For given PA positions $\boldsymbol{\psi}$ and $\boldsymbol{\phi}$, define the communication channel matrix as
	$\mathbf H_c=[\mathbf h_{c,1},\ldots,\mathbf h_{c,K_C}]^{\top}\in\mathbb C^{K_C\times M}$.
	Assuming $\mathbf H_c$ has full row rank, the ZF precoding matrix is
	\begin{equation}
		\mathbf V_{\rm ZF}=\mathbf H_c^{\mathrm H}
		(\mathbf H_c\mathbf H_c^{\mathrm H})^{-1},
	\end{equation}
	where $\mathbf H_c\mathbf V_{\rm ZF}=\mathbf I_{K_C}$.
	The communication beamforming matrix is constructed as
	$\mathbf W_C=\mathbf V_{\rm ZF}\operatorname{diag}
	(\sqrt{p_1},\ldots,\sqrt{p_{K_C}})$,
	where $p_k$ denotes the received signal power of CU $k$ under ZF.
	Accordingly, the minimum power allocation satisfying the rate constraints is
	\begin{equation}
		p_k^{\min}=(2^{\Gamma_k}-1)\sigma_c^2.
	\end{equation}
	Let $a_k=\|\mathbf v_k\|^2$, where $\mathbf v_k$ is the $k$-th column of $\mathbf V_{\rm ZF}$.
	The remaining transmit power is
	\begin{equation}
		P_{\rm rem}=P_t-\sum_{k=1}^{K_C}a_kp_k^{\min}.
	\end{equation}
	If $P_{\rm rem}<0$, the current PA configuration is regarded as infeasible for the ZF-based initialization. For $P_{\rm rem}\geq 0$, we construct up to $K_C+2$ candidate beamforming matrices by allocating the remaining power as follows: 1) Allocate all remaining power to the $j$-th communication beam, yielding $p_j=p_j^{\min}+P_{\rm rem}/a_j$, for $j=1,\ldots,K_C$, while keeping $p_k=p_k^{\min}$ for $k\neq j$; 2) Equally allocate the remaining transmit power among all communication beams, yielding $p_k=p_k^{\min}+P_{\rm rem}/(K_Ca_k)$; 3) Allocate the remaining power to a dedicated sensing beam $\mathbf w_S=\sqrt{P_{\rm rem}}\mathbf u_S$, while setting $p_k=p_k^{\min}$.
	
	For the third scheme, the sensing beam is designed within the null space of $\mathbf H_c$ to avoid communication interference. Let $\mathbf U_0$ be an orthonormal basis of $\operatorname{null}(\mathbf H_c)$, such that
	$\mathbf u_S=\mathbf U_0\mathbf q$ with $\|\mathbf q\|=1$.
	To determine the sensing direction, we maximize its contribution to $\operatorname{Tr}(\mathbf F_{\xi\xi})$ as a surrogate sensing metric, yielding
	\begin{equation}
		\mathbf q^{\star}=\mathbf v_{\max}(\mathbf G_S),
	\end{equation}
	where $\mathbf v_{\max}(\cdot)$ denotes the principal eigenvector and
	\begin{equation}
		\mathbf G_S=\sum_{k=1}^{K_T}\sum_{\nu\in\{x,y\}}
		(\dot{\mathbf H}_{\nu_{t,k}}\mathbf U_0)^{\mathrm H}
		(\dot{\mathbf H}_{\nu_{t,k}}\mathbf U_0).
	\end{equation}
	Here, $\dot{\mathbf H}_{\nu_{t,k}}$ denotes the derivative of the target echo channel with respect to the corresponding target coordinate, as defined in Appendix~A.
	The sensing direction is then $\mathbf u_S=\mathbf U_0\mathbf q^{\star}$.
	If $\operatorname{null}(\mathbf H_c)=\{\mathbf 0\}$ and $P_{\rm rem}>0$, the third scheme is omitted.
	
	Finally, each candidate beamforming matrix is formed as
	$\mathbf W=[\mathbf W_C,\mathbf w_S]$,
	where $\mathbf w_S=\mathbf 0$ for the first $K_C+1$ schemes.
	The candidate achieving the minimum CRLB is selected as
	\begin{equation}
		\mathbf W^{\star}_{\rm ZF}
		=\underset{\mathbf W\in\mathcal W_{\rm ZF}}{\arg\min}
		\operatorname{Tr}(\mathbf F_{\rm eq}^{-1}),
	\end{equation}
	where $\mathcal W_{\rm ZF}$ contains the feasible candidate beamforming matrices.

	\bibliographystyle{IEEEtran}
	\bibliography{Refs}{}

\begin{thebibliography}{10}
\providecommand{\url}[1]{#1}
\csname url@samestyle\endcsname
\providecommand{\newblock}{\relax}
\providecommand{\bibinfo}[2]{#2}
\providecommand{\BIBentrySTDinterwordspacing}{\spaceskip=0pt\relax}
\providecommand{\BIBentryALTinterwordstretchfactor}{4}
\providecommand{\BIBentryALTinterwordspacing}{\spaceskip=\fontdimen2\font plus
\BIBentryALTinterwordstretchfactor\fontdimen3\font minus
  \fontdimen4\font\relax}
\providecommand{\BIBforeignlanguage}[2]{{%
\expandafter\ifx\csname l@#1\endcsname\relax
\typeout{** WARNING: IEEEtran.bst: No hyphenation pattern has been}%
\typeout{** loaded for the language `#1'. Using the pattern for}%
\typeout{** the default language instead.}%
\else
\language=\csname l@#1\endcsname
\fi
#2}}
\providecommand{\BIBdecl}{\relax}
\BIBdecl

\bibitem{mimo}
Z.~Wang \emph{et~al.}, ``{A tutorial on extremely large-scale MIMO for 6G:
  Fundamentals, signal processing, and applications},'' \emph{IEEE Commun.
  Surv. Tutor.}, vol.~26, no.~3, pp. 1560--1605, 2024.

\bibitem{ra}
W.~Ma \emph{et~al.}, ``{A survey on reconfigurable and movable antennas for
  wireless communications and sensing},'' \emph{IEEE Commun. Surv. Tutor.},
  vol.~28, pp. 4842--4882, 2026.

\bibitem{ris}
Y.~Liu \emph{et~al.}, ``{Reconfigurable intelligent surfaces: Principles and
  opportunitie}s,'' \emph{IEEE Commun. Surv. Tutor.}, vol.~23, no.~3, pp.
  1546--1577, 2021.

\bibitem{fas}
T.~Wu \emph{et~al.}, ``{Fluid antenna systems enabling 6G: Principles,
  applications, and research directions},'' \emph{IEEE Wirel. Commun.}, pp.
  1--9, 2025.

\bibitem{ma}
L.~Zhu, W.~Ma, and R.~Zhang, ``{Movable antennas for wireless communication:
  Opportunities and challenges},'' \emph{IEEE Commun. Mag.}, vol.~62, no.~6,
  pp. 114--120, 2024.

\bibitem{pa1}
Z.~Yang \emph{et~al.}, ``{Pinching antennas: Principles, applications and
  challenges},'' \emph{IEEE Wirel. Commun.}, pp. 1--10, 2025.

\bibitem{pa2}
Y.~Liu \emph{et~al.}, ``{Pinching-antenna systems: Architecture designs,
  opportunities, and outlook},'' \emph{IEEE Commun. Mag.}, vol.~64, no.~1, pp.
  190--196, 2026.

\bibitem{pa3}
Z.~Ding, R.~Schober, and H.~Vincent~Poor, ``{Flexible-antenna systems: A
  pinching-antenna perspective},'' \emph{IEEE Trans. Commun.}, vol.~73, no.~10,
  pp. 9236--9253, 2025.

\bibitem{pa4}
H.~O.~Y. Suzuki and K.~Kawai, ``{Pinching antenna: Using a dielectric waveguide
  as an antenna},'' \emph{NTT DOCOMO Technical J}, vol.~23, no.~3, pp. 5--12,
  2022.

\bibitem{pa5}
Z.~Wang, C.~Ouyang, X.~Mu, Y.~Liu, and Z.~Ding, ``{Modeling and beamforming
  optimization for pinching-antenna systems},'' \emph{IEEE Trans. Commun.},
  vol.~73, no.~12, pp. 13\,904--13\,919, 2025.

\bibitem{pa6}
M.~Zeng, J.~Wang, G.~Zhou, F.~Fang, and X.~Wang, ``{Energy-efficient design for
  downlink pinching-antenna systems with QoS guarantee},'' \emph{IEEE Trans.
  Veh. Technol.}, vol.~75, no.~2, pp. 3503--3507, 2026.

\bibitem{pa7}
M.~Sun, C.~Ouyang, S.~Wu, and Y.~Liu, ``Robust beamforming for pinching-antenna
  systems,'' \emph{IEEE Trans. Veh. Technol.}, vol.~75, no.~7, pp.
  15\,073--15\,078, 2026.

\bibitem{pa8}
{M. Sun, C. Ouyang, S. Wu, and Y. Liu}, ``{Multiuser beamforming for
  pinching-antenna systems: An element-wise optimization framework},''
  \emph{IEEE Trans. Wireless Commun.}, vol.~25, pp. 6538--6552, 2026.

\bibitem{pa4sen1}
Z.~Wang, C.~Ouyang, Y.~Liu, and A.~Nallanathan, ``{Wireless sensing via
  pinching-antenna systems},'' \emph{IEEE Wirel. Commun. Lett.}, vol.~14,
  no.~11, pp. 3475--3479, 2025.

\bibitem{pa4isac1}
Y.~Qin, Y.~Fu, and H.~Zhang, ``{Joint antenna position and transmit power
  optimization for pinching antenna-assisted ISAC systems},'' \emph{IEEE Wirel.
  Commun. Lett.}, vol.~14, no.~11, pp. 3535--3539, 2025.

\bibitem{pa4isac2}
W.~Mao \emph{et~al.}, ``{Multi-waveguide pinching antennas for ISAC},''
  \emph{IEEE Trans. Wireless Commun.}, vol.~25, pp. 5846--5858, 2026.

\bibitem{pa4isac3}
H.~Li \emph{et~al.}, ``{Pinching antenna systems for integrated sensing and
  communications},'' \emph{IEEE Trans. Wireless Commun.}, vol.~25, pp.
  13\,416--13\,429, 2026.

\bibitem{pa4isac4}
D.~Bozanis, V.~K. Papanikolaou, S.~A. Tegos, and G.~K. Karagiannidis,
  ``{Cramér-Rao bounds for integrated sensing and communications in
  pinching-antenna systems},'' in \emph{Proc. IEEE Int. Symp. Pers. Indoor Mob.
  Radio Commun.}, 2025, pp. 1--6.

\bibitem{swan1}
C.~Ouyang, H.~Jiang, Z.~Wang, Y.~Liu, and Z.~Ding, ``{Uplink and downlink
  communications in segmented waveguide-enabled pinching-antenna systems
  (SWANs)},'' \emph{IEEE Trans. Commun.}, vol.~74, pp. 3688--3703, 2026.

\bibitem{swan2}
S.~Gu, H.~Jiang, C.~Ouyang, Y.~Liu, and D.~I. Kim, ``Sum-rate maximization for
  uplink segmented waveguide-enabled pinching-antenna systems,'' \emph{IEEE
  Wirel. Commun. Lett.}, vol.~15, pp. 3876--3880, 2026.

\bibitem{swan3}
Q.~Gao, R.~Zhong, H.~Shin, and Y.~Liu, ``{Integrated sensing and communication
  for segmented waveguide-enabled pinching antenna systems},'' \emph{IEEE
  Trans. Veh. Technol.}, early access, 2026.

\bibitem{swan4}
H.~Jiang \emph{et~al.}, ``{Exploiting segmented waveguide-enabled
  pinching-antenna systems (SWANs) in ISAC},'' in \emph{Proc. IEEE INFOCOM},
  2026, pp. 1--6.

\bibitem{signal}
X.~Liu, T.~Huang, N.~Shlezinger, Y.~Liu, J.~Zhou, and Y.~C. Eldar, ``{Joint
  transmit beamforming for multiuser MIMO communications and MIMO rada}r,''
  \emph{IEEE Trans. Signal Process.}, vol.~68, pp. 3929--3944, 2020.

\bibitem{fim}
H.~Hua, J.~Xu, and Y.~C. Eldar, ``{Near-field 3D localization via MIMO radar:
  Cramér-rao bound analysis and estimator design},'' \emph{IEEE Trans. Signal
  Process.}, vol.~72, pp. 3879--3895, 2024.

\bibitem{CE}
J.~Xiao, J.~Wang, and Y.~Liu, ``{Channel estimation for pinching-antenna
  systems (PASS)},'' \emph{IEEE Commun. Lett.}, vol.~29, no.~8, pp. 1789--1793,
  2025.

\bibitem{rmo}
N.~Boumal, \emph{{An Introduction to Optimization on Smooth Manifolds}}.\hskip
  1em plus 0.5em minus 0.4em\relax Cambridge University Press, 2023.

\bibitem{penalty}
J.~Nocedal and S.~J. Wright, \emph{{Numerical Optimization}}.\hskip 1em plus
  0.5em minus 0.4em\relax Springer New York, NY, 2006.

\bibitem{rep}
C.~Liu and N.~Boumal, ``{Simple algorithms for optimization on Riemannian
  manifolds with constraints},'' \emph{Appl. Math. Optim.}, vol.~82, no.~3, pp.
  949--981, 2020.

\bibitem{lq}
M.~c. Pinar and S.~A. Zenios, ``On smoothing exact penalty functions for convex
  constrained optimization,'' \emph{SIAM J. Optim.}, vol.~4, no.~3, pp.
  486--511, 1994.

\bibitem{CM}
A.~Hjorungnes and D.~Gesbert, ``{Complex-valued matrix differentiation:
  Techniques and key results},'' \emph{IEEE Trans. Signal Process.}, vol.~55,
  no.~6, pp. 2740--2746, 2007.

\bibitem{rbfgs}
W.~Huang, P.-A. Absil, and K.~A. Gallivan, ``{A Riemannian BFGS method for
  nonconvex optimization problems},'' in \emph{Numerical Mathematics and
  Advanced Applications ENUMATH 2015}.\hskip 1em plus 0.5em minus 0.4em\relax
  Springer, 2016, pp. 627--634.

\bibitem{lbfgs}
W.~Huang, K.~A. Gallivan, and P.-A. Absil, ``{A Broyden class of quasi-Newton
  methods for Riemannian optimization},'' \emph{SIAM Journal on Optimization},
  vol.~25, no.~3, pp. 1660--1685, 2015.

\end{thebibliography}
	
\end{document}